\RequirePackage{fix-cm}
\documentclass{svjour3}                     

\smartqed  
\usepackage{amsmath}
\usepackage{graphicx}
\usepackage[T1]{fontenc}
\usepackage[utf8,latin1]{inputenc}
\usepackage{epsfig}
\usepackage{amssymb}
\usepackage{natbib}        
\usepackage{xspace}

\usepackage{url}
\usepackage{color}
\usepackage{bm}
\usepackage{sgl-commands}

\journalname{Space Science Reviews}
\newcommand{\sn}{SN\xspace}
\newcommand{\sne}{SNe\xspace}
\newcommand{\snia}{SN~Ia\xspace}
\newcommand{\sneia}{SNe~Ia\xspace}

\newcommand{\dt}{D_{\Delta t}} 
\newcommand{\OM}{\Omega_{\rm m}} 
\newcommand{\kmsMpc}{{\rm km}\,{\rm s}^{-1}\,{\rm Mpc}^{-1}}

\begin{document}


\title{Strong gravitational lensing and microlensing of supernovae}


\titlerunning{Strong and microlensing of supernovae}        

\author{Sherry H.~Suyu$^{1,2,3}$ \and Ariel Goobar$^4$ \and Thomas Collett$^5$ \and Anupreeta More$^{6,7}$ \and Giorgos Vernardos$^{8}$ }

\authorrunning{Suyu, Goobar, Collett et al.} 

\institute{Sherry H.~Suyu \at \email{suyu@mpa-garching.mpg.de}
\and 
Ariel Goobar \at \email{ariel@fysik.su.se}
\and
Thomas Collett \at \email{thomas.collett@port.ac.uk} 
\and 
Anupreeta More \at \email{anupreeta@iucaa.in}
\and 
Giorgos Vernardos \at \email{georgios.vernardos@epfl.ch} 
\\
\\
$^1$Technical University of Munich, TUM School of Natural Sciences, Department of Physics, James-Franck-Stra\ss{}e~1, 85748 Garching, Germany\\ 
$^2$Max-Planck-Institut f{\"u}r Astrophysik, Karl-Schwarzschild-Str.~1, 85748 Garching, Germany\\
$^3$Institute of Astronomy and Astrophysics, Academia Sinica, 11F of ASMAB, No.1, Section 4, Roosevelt Road, Taipei 10617, Taiwan\\
$^4$The Oskar Klein Centre, Department of Physics, Stockholm University, Albanova University Center, SE-106 91 Stockholm, Sweden\\
$^5$Institute of Cosmology and Gravitation, University of Portsmouth, Dennis Sciama Building, Burnaby Road, Portsmouth, PO1 3FX, UK \\
$^6$The Inter-University Centre for Astronomy and Astrophysics (IUCAA), Post Bag 4, Ganeshkhind, Pune 411007, India\\
$^7$Kavli Institute for the Physics and Mathematics of the Universe (IPMU), 5-1-5 Kashiwanoha, Kashiwa-shi, Chiba 277-8583, Japan\\
$^8$Institute of Physics, Laboratory of Astrophysics, Ecole Polytechnique F\'{e}d\'{e}rale de Lausanne (EPFL), Observatoire de Sauverny, 1290 Versoix, Switzerland}

\date{Draft: \today / Received: date / Accepted: date}



\maketitle

\begin{abstract}
Strong gravitational lensing and microlensing of supernovae (SNe) are emerging as a new probe of cosmology and astrophysics in recent years.  We provide an overview of this nascent research field, starting with a summary of the first discoveries of strongly lensed SNe.  We describe the use of the time delays between multiple SN images as a way to measure cosmological distances and thus constrain cosmological parameters, particularly the Hubble constant, whose value is currently under heated debates.  New methods for measuring the time delays in lensed SNe have been developed, and the sample of lensed SNe from the upcoming Rubin Observatory Legacy Survey of Space and Time (LSST) is expected to provide competitive cosmological constraints.  Lensed SNe are also powerful astrophysical probes.  We review the usage of lensed SNe to constrain SN progenitors, acquire high-z SN spectra through lensing magnifications, infer SN sizes via microlensing, and measure properties of dust in galaxies. The current challenge in the field is the rarity and difficulty in finding lensed SNe.  We describe various methods and ongoing efforts to find these spectacular explosions, forecast the properties of the expected sample of lensed SNe from upcoming surveys particularly the LSST, and summarize the observational follow-up requirements to enable the various scientific studies.  We anticipate the upcoming years to be exciting with a boom in lensed SN discoveries.

\end{abstract}


\tableofcontents

\label{chapter1}

\section{Brief history}
\label{sec:LSNe:history}

In an insightful and pioneering publication, \citet{1964MNRAS.128..307R} pointed out that supernovae (\sne) would be particularly interesting sources for studies involving strong gravitational lensing. This was arguably among the biggest early milestones in the field of gravitational lensing, following the realisation by \cite{PhysRev.51.290} that the scenario, first proposed by \citeauthor{1936Sci....84..506E} in 1936, had realistic applications in extra-galactic astronomy. Refsdal pointed out that the measurement of the time delay between the arrival of \sn images could be used to infer the Hubble constant ($H_0$). 

\subsection{History of searches for lensed SNe behind clusters}
\label{sec:LSNe:history:search}

The idea to use ``gravitational telescopes'', i.e., known lensing galaxy clusters, to boost the faint signals from distant supernovae started to get traction about thirty years ago \citep{1988ApJ...335L...9K,2000MNRAS.319..549S,2003A&A...405..859G}. 
Since the lensing magnification boosts the signal from the faint distant source behind the lens by a factor of $\mu$, but leaves the dominant foreground sky noise unaffected, the signal-to-noise ratio scales as $SNR \propto \mu \sqrt{t}$, hence leading to a gain factor  $\mu^2$ in exposure length. 
However, the solid angle at the source planes shrinks by a factor $\mu$ behind the lens. Thus, the net gain/loss of searching for supernovae behind massive clusters is therefore a non-trivial combination of field of view, limiting depth, and supernova luminosity functions. Although not spectroscopically confirmed, several \sn candidates were eventually found in cadenced observations in the Near-IR at the Very Large Telescopes (VLT) of the European Southern Observatory (ESO) \citep{2009A&A...507...61S,2009A&A...507...71G}, including a $z=1.7$ core-collapse \sn\ behind Abell 1689, with a lens model inferred magnification $\mu=4.3 \pm 0.3$ \citep{2011ApJ...742L...7A}. In spite of the limitations of the survey, \citet{2016A&A...594A..54P} demonstrated the power of gravitational lensing to place meaningful limits on the rate of core-collapse supernovae at very high redshifts.

\subsection{Discoveries!}   
\label{sec:LSNe:history:discovery}
 
PS1-10afx was first reported by \citet{2013ApJ...767..162C} as an unusual superluminous \sn. However, shortly thereafter, \citet{2013ApJ...768L..20Q} showed that it was a perfect match to a lensed SN Ia at redshift $z = 1.388$, with a large amplification, $\mu \sim 30$. \citet{Quimby+2014} eventually also identified a foreground lens, at $z = 1.117$. Even if the putative lensing object was too close to the SN, it was consistent with lensing models and existing data. Since the lensed \snia classification only gained acceptance four years after the explosion, further investigation of the lensing nature of the SN or its type was not possible. Moreover, a subsequent single-band HST imaging (PI: Chornock) proved insufficient to verify the presence of any lensed images of the SN host galaxy. 

\begin{figure*}
    \centering
    \includegraphics[width=\textwidth]{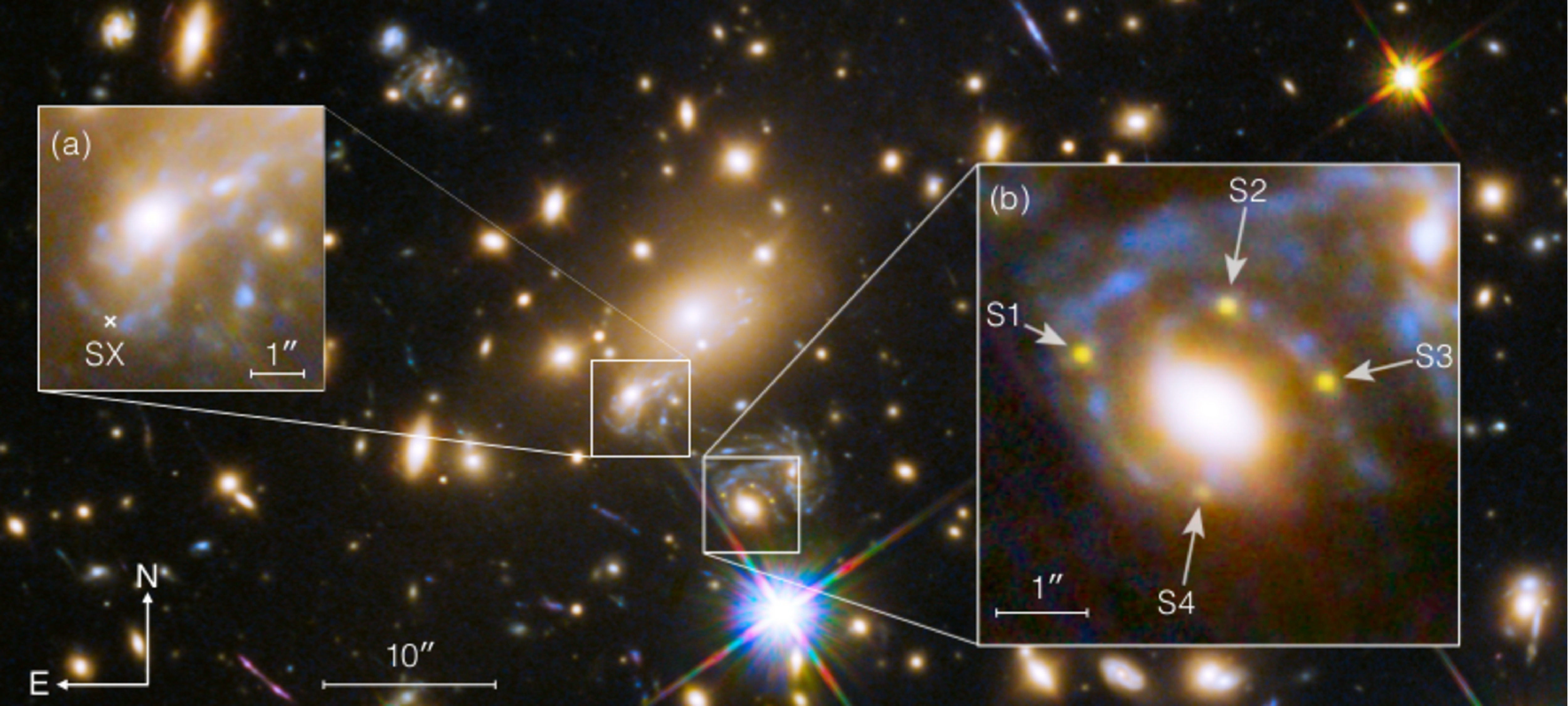}
    \caption{Hubble Space Telescope image of SN Refsdal, the first strongly lensed SN system with spatially resolved images.  Inset (a) shows the image SX that was detected in December 2015 \citep{Kelly+2016}, and inset (b) shows the multiple images S1, S2, S3 and S4 that were first discovered in November 2014 \citep{2015Sci...347.1123K}.  There is another image SY located in the northern most image of the spiral host galaxy, next to the top-right corner of the inset (a).  Image taken from \citet{Grillo+2018}.  Original image credit: NASA, ESA/Hubble.  }
    \label{fig:SNRefsdal}
\end{figure*}

The first detection of multiple images from a supernova came with SN Refsdal  at redshift $z=1.49$ \citep{2015Sci...347.1123K}, found in Hubble Space Telescope (HST) surveys of the massive lensing galaxy cluster MACS J1149.6+2223 at $z = 0.54$. Based on both light curve shapes and spectroscopy, SN Refsdal was classified as a Type II SN, resembling the iconic SN 1987A in the LMC \citep{2016ApJ...831..205K}. This is a peculiar class of faint \sne, quite rare in the local universe. 
At the time of discovery, 4 multiple images of the SN were visible with HST (images S1-S4, see Fig.~\ref{fig:SNRefsdal} taken from \citealt{Grillo+2018}), in a cross configuration around one of the foreground cluster galaxies. There are two other images of the background spiral galaxy hosting the SN, labelled as SX (inset (a) in Fig.~\ref{fig:SNRefsdal}) and SY (the northern-most image, just outside the top-right corner of inset (a)). 

Using mass models of the cluster, the discovery team \citep{2015Sci...347.1123K} predicted that SY happened before S1-S4, whereas SX would appear after S1-S4.  However, the time of SX reappearance was uncertain from their model, ranging from 1 to 10 years.  To refine the delay prediction, \citet{Grillo+2016} used the Multi Unit Spectroscopic Explorer \citep[MUSE;][]{Bacon+2010} on ESO's VLT to obtain spectroscopic observations of the field.  The spectroscopic redshift measurements helped to separate foreground lenses from background sources.  This spectroscopic data set was shared with multiple modelling teams, who attempted to predict the reappearance of image SX using the new observational data. Teams predicted their delays before the reappearance \citep{Treu+2016, Jauzac+2016, Grillo+2016, Kawamata+2016}, providing a true blind test of the predictions and the modelling capabilities. Most teams predicted a short time delay of SX with respect to S1, within a year's time.  \citet{Kelly+2016} detected the reappearance of SX in December 2015, and the resulting constraint on the time delay and magnification of SX agreed well with the predictions from two teams \citep{Grillo+2016, Kawamata+2016}.  

Through the monitoring of the multiple images of SN Refsdal, \citet{Rodney+2016} have measured the time delays and magnification ratios among images S1, S2, S3 and S4.  The detection of the reappearance of image SX provided an estimate of its time delay with respect to the other images \citep{Kelly+2016}, and the time-delay measurement from subsequent monitoring is forthcoming (P.~Kelly, private communications).  The SX time delay is expected to be precise with uncertainties of at most a few percent given its long delay of $\sim$1 year, providing a great opportunity of measuring the Hubble constant with gravitational lens time delays \citep[e.g.,][Kelly et al. in prep.]{Grillo+2018, Grillo+2020}.

\begin{figure*}
    \centering
    \includegraphics[width=\textwidth]{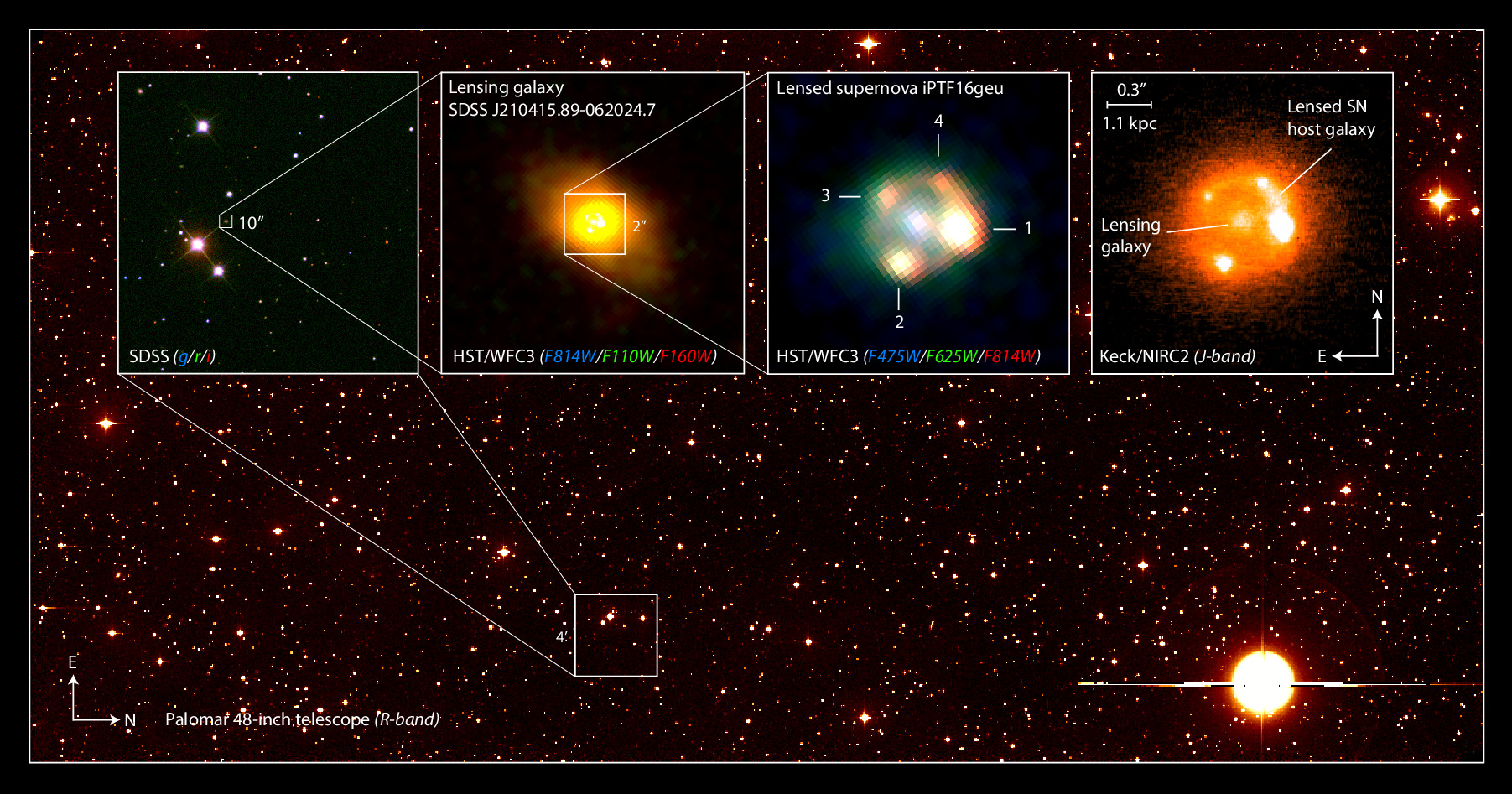}
    \caption{Wide-field image from iPTF showing the portion of the CCD camera where iPTF16geu was found in modest seeing ($2^{\prime\prime}$). The insets show how the strong lensing nature, a quadruple lens with an Einstein radius of only $0.3^{\prime\prime}$, could be verified using HST imaging in the optical, and laser-guide-star adaptive-optics imaging from Keck in the Near-IR. Image credit: J.~Johansson.}
    \label{fig:iptf16geu}
\end{figure*}

The first spatially resolved multiply-imaged Type Ia supernova, iPTF16geu, was detected by the intermediate Palomar Transient Factory \citep{2017Sci...356..291G}. As shown in Fig.~\ref{fig:iptf16geu}, the ground-based imager at iPTF could not spatially resolve the very compact system with Einstein radius $\tae =0.3\arcsec$. However, because of the ``standard candle'' nature of \sneia, it became clear that strong lensing was the most likely explanation, since the SN was more than 30 standard deviations brighter than the expectations for a \snia at its measured redshift, $z=0.409$. The spectra used to classify the supernova showed spectral lines from both the host galaxy at the same redshift, but also the deflecting galaxy at $z=0.216$. Thanks to multi-band follow-up with HST, an accurate (model independent) measurement of the magnification 
was made, $\mu=67.8^{+2.6}_{-2.9}$ \citep{2020MNRAS.491.2639D}, after correction for non-negligible extinction by dust in both the host and lens galaxies. The time delays between the SN images for this system were very small, about a day or less \citep{More+2017,2020MNRAS.491.2639D}. The flux ratios between the supernova images (see Fig.~\ref{fig:iptf16geu}) were not consistent with expectations from lensing, hinting contributions from stellar microlensing \citep{More+2017}. This was further confirmed after differential dust extinction from within the lensing galaxy was also accounted for \citep{2020MNRAS.496.3270M}.

In 2021, another strongly lensed SN, SN Requiem, was discovered in archival HST 
imaging of the galaxy cluster MACSJ0138.0$-$2155 by \citet{Rodney+2021}.  A bright quiescent and evolved elliptical galaxy at redshift $z=1.95$ is lensed into giant arcs by the galaxy cluster, making this system one of the brightest NIR objects \citep[e.g.,][]{Newman+2018}. Near the giant arcs are three point sources that were present in the single-epoch image in 2016 \citep{Newman+2018}, but absent in the 2019 HST image (HST Program - REQUIEM; HST-GO-15663; PI:Akhshik).  These 3 point sources are identified as multiple images of a SN hosted in the $z=1.95$ galaxy, and the SN is likely to be of Type Ia.  Based on the differences in the colour of the 3 SN images and SN Ia templates, the time delays could be estimated from the single-epoch observation in 2016, although with large uncertainties. The predicted time delays suggest that the fourth image of SN Requiem will appear in the future (circa 2037). 

\begin{figure*}
    \centering
    \includegraphics[width=\textwidth]{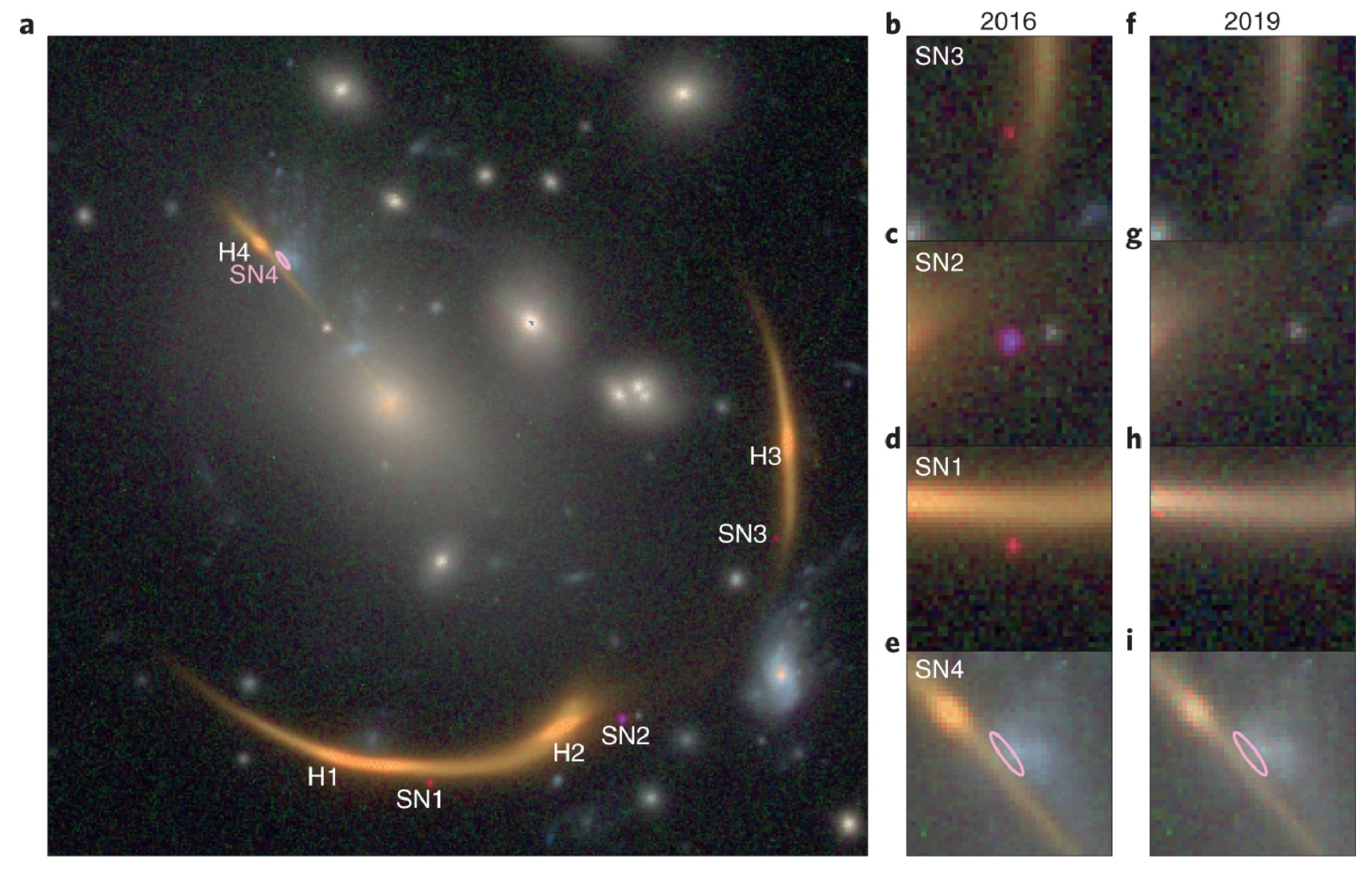}
    \caption{Panel a): Color image of SN Requiem showing arc-like images of the distant host galaxy (H1-H4), the three SN images (SN1-SN3) and an ellipse pointing out the expected location of SN4. Panels b-i): zoomed-in regions around the SN locations from the data taken in 2016 (b-e) and 2019 (f-i). Figure taken from \citet{Rodney+2021}.}
    \label{fig:snrequiem}
\end{figure*}

Recently, \citet{Goobar+2022} discovered another strongly lensed SNe Ia, named Supernova Zwicky (a.k.a.~SN 2022qmx), in the Zwicky Transient Facility \citep[ZTF;][]{Bellm+2019}.  The discovery of this system is similar to the case of iPTF16geu -- the multiple lensed SN Ia images are closely separated (with an Einstein radius of $\tae \sim 0.17\arcsec$) and not spatially resolved by ZTF, thus resulting in a brightness through lensing magnification that is substantially higher than expected for a SN Ia.  \citet{Pierel+2022} obtained HST imaging of this system, showing clearly the four multiple images of the SN Ia in a symmetric configuration.  Based on the single-epoch photometry of the SN Ia images, \citet{Pierel+2022} measured short time delays of $<1$\,day between the multiple images, which are consistent with the predictions from their multiple lens mass models obtained using different lensing software.  Both \citet{Goobar+2022} and \citet{Pierel+2022} find anomalous flux ratios of the SN images compared to the smooth  mass model predictions, indicating the presence of dust, millilensing and/or microlensing. 

The fifth system of spatially-resolved lensed SN is reported very recently by \citet{Chen+2022}.  A SN was lensed by one of the galaxies in the galaxy cluster Abell 370 in 2010 into three multiple images, and \citet{Chen+2022} discovered it in archival HST image, similar to the way that SN Resfdal was found.  Based on their cluster lens mass modeling, \citet{Chen+2022} estimated the rest-frame age of the SN for each of the three SN images, with the youngest one being a mere $\sim 6$ hours after explosion.  Using the early-phase light curve obtained in the single-epoch HST observations with multiple filters, \citet{Chen+2022} classified the SN as a core-collapse SN and measured its pre-explosion radius of $\sim500 M_{\odot}$, consistent with a red supergiant. The estimated photometric redshift of the SN host galaxy is $\sim3$.

These first few strongly lensed SNe have opened a new window to probe cosmology and supernovae through these spectacular phenomena.  Two reviews \citep{Oguri2019, Liao+2022} on gravitationally lensed transients have illustrated the exciting utilities of lensed transients.  Compared to these two reviews, we focus and dive deeper into the specific case of strongly lensed supernovae. In Section \ref{sec:LSNe:cosmo_probe}, we review the potential of lensed SNe as a cosmological probe. In Section \ref{sec:LSNe:astro_probe}, we describe the use of lensed SNe as an astrophysical probe, such as constraining the SN progenitors and dust in galaxies.  In Section \ref{sec:LSNe:search}, we present various searches of lensed SNe and the expected rates in current and upcoming surveys.  We summarise in Section \ref{sec:LSNe:summary}.

\section{Cosmological probe}
\label{sec:LSNe:cosmo_probe}

While the idea of using lensed SNe for measuring the Hubble constant dates back to \citet{1964MNRAS.128..307R}, this time-delay method was first realised with lensed quasars that were discovered decades before lensed SNe, starting with the first lensed quasar system found by \citet{Walsh+1979}.  The use of lensed quasars as a cosmological probe is reviewed by \citet{Birrer+2022} in the same series \citep[see also, e.g.,][]{TreuMarshall2016, Suyu+2018, Treu+2022}.  We briefly summarise the approach of time-delay cosmography in Section \ref{sec:LSNe:cosmo_probe:candle}, and focus on SNe as lensed background sources, which are ``standardisable candles'' and a relative distance indicator. In Section \ref{sec:LSNe:cosmo_probe:td_spectro}, we describe approaches to measure time delays of lensed SNe, particularly novel ones through spectroscopic observations of SNe, which complement conventional techniques using photometric light curves.  In Section \ref{sec:LSNe:cosmo_probe:present_future}, we show the present cosmological constraints and future forecasts from lensed SNe.  

\subsection{Jackpot: two cosmological probes in one} 
\label{sec:LSNe:cosmo_probe:candle}

We describe the various distance measurements that we can obtain from strongly lensed SNe, through the lensing time delays and the SN light curves.  We briefly summarise the determination of the lensing distances that are covered in detail in the parallel reviews by, e.g., \citet{Birrer+2022},
and focus particularly on the new aspects that SNe bring.

The expression for the time delay between images $i$ and $j$ of a lensed SN is 
\begin{equation}
\label{eq:cosmo:timedelay}
    \Delta t_{ij} = \frac{\dt}{c} \Delta \tau_{ij},
\end{equation}
where $\dt$ is the time-delay distance, $c$ is the speed of light, and $\tau$ is the Fermat potential.  The time-delay distance \citep[e.g.,][]{1964MNRAS.128..307R, Suyu+2010} is defined by 
\begin{equation}
\label{eq:cosmo:Dt}
\dt \equiv (1+\zd) \frac{\dd \ds}{\dds},
\end{equation}
where $\zd$ is the deflector (lens) redshift, and $\dd$, $\ds$, and $\dds$ are the angular diameter distance to the deflector, to the source, and between the deflector and the source, respectively.  The Fermat potential difference between two SN image positions $\tang_i$ and $\tang_j$ (with corresponding source position $\bang$) is
\begin{equation}
    \label{eq:cosmo:fermatpotdiff}
\Delta \tau_{ij} = \tau(\tang_i;\bang) - \tau(\tang_j;\bang),
\end{equation}
where the Fermat potential is 
\begin{equation}
    \label{eq:cosmo:fermatpot}
    \tau(\tang;\bang) = \frac{1}{2}(\tang-\bang)^2 - \psi(\tang),
\end{equation}
and $\psi(\tang)$ is the lens potential.  The dimensionless surface mass density $\kappa(\tang)$, a.k.a. lensing convergence, is related to the lens potential via the Poisson equation,
\begin{equation}
    \label{eq:cosmo:kappa}
    2\kappa(\tang) = \nabla^2\psi(\tang)
\end{equation}

From Equation (\ref{eq:cosmo:timedelay}), we see that by measuring the time delays $\Delta t_{ij}$ and modelling the deflection and line-of-sight mass distributions to get $\Delta \tau_{ij}$, we can infer the time-delay distance $\dt$.  Since $\dt$ is a combination of three angular diameter distances (Equation \ref{eq:cosmo:Dt}) and is thus proportional to $H_0^{-1}$, time-delay lenses allow us to measure directly $H_0$ with weak dependence on other cosmological parameters.  

When the lens galaxy stellar velocity dispersion is measured, then the combination of time delays, velocity dispersion and high-resolution imaging of the lens system allows $\dd$ to be measured \citep[e.g.,][]{ParaficzHjorth2009, Jee+2015, Jee+2019}, in addition to $\dt$.  The joint constraint of $\dt$ and $\dd$ provides even more leverage on determining cosmological parameters  \citep[e.g.,][]{Jee+2016, Yildirim+2020}.

It is clear from Equation (\ref{eq:cosmo:fermatpot}) that in order to constrain $H_0$ with strong lens time delays, one needs to know both the 2D lens potential and the unlensed source position, neither of which are directly observable. The use of lens modelling is  required to infer these quantities. However, strong lens models are subject to degeneracies, which are a major source of uncertainty for time-delay cosmography \citep{SchneiderSluse2014}. The largest problem is the mass-sheet degeneracy: re-scaling any model of the lensing convergence $\kappa$ and adding a constant sheet of surface mass density leaves the predicted images unchanged, but alters the predicted time delay between the images \citep{Falco+1985}. 
Breaking the mass-sheet degeneracy is therefore necessary to constrain $H_0$ from any lens with observations of image positions and time delays. In order to break the mass-sheet degeneracy, additional information is required, either non-lensing information (e.g., the dynamical mass of the system), the presence of sources at multiple redshifts \citep[though see][]{Schneider2014} or knowledge of the intrinsic luminosity or size of the unlensed source \citep{KolattBartelmann1998}. It is on this final point -- a known luminosity of the source -- that strongly lensed SNe shine as a powerful potential cosmological probe. 

In this context, Type Ia SNe play a special role. Thanks to the homogeneity of their luminosity, these thermonuclear explosions could be used as sharp distance estimators to measure the accelerated expansion of the universe \citep{1998AJ....116.1009R,1999ApJ...517..565P}, leading to the discovery of dark energy. State-of-the-art \snia cosmological surveys have demonstrated that the intrinsic luminosity of \sneia only varies by about 0.1 mag \citep{2014A&A...568A..22B,2018ApJ...859..101S}. \citet{Birrer+2021} have shown the great benefit of these sharp distance indicators to break the mass-sheet degeneracy to improve the accuracy of the measurements of the Hubble constant from time delays.

\begin{figure*}
    \centering
    \includegraphics[width=0.25\textwidth]{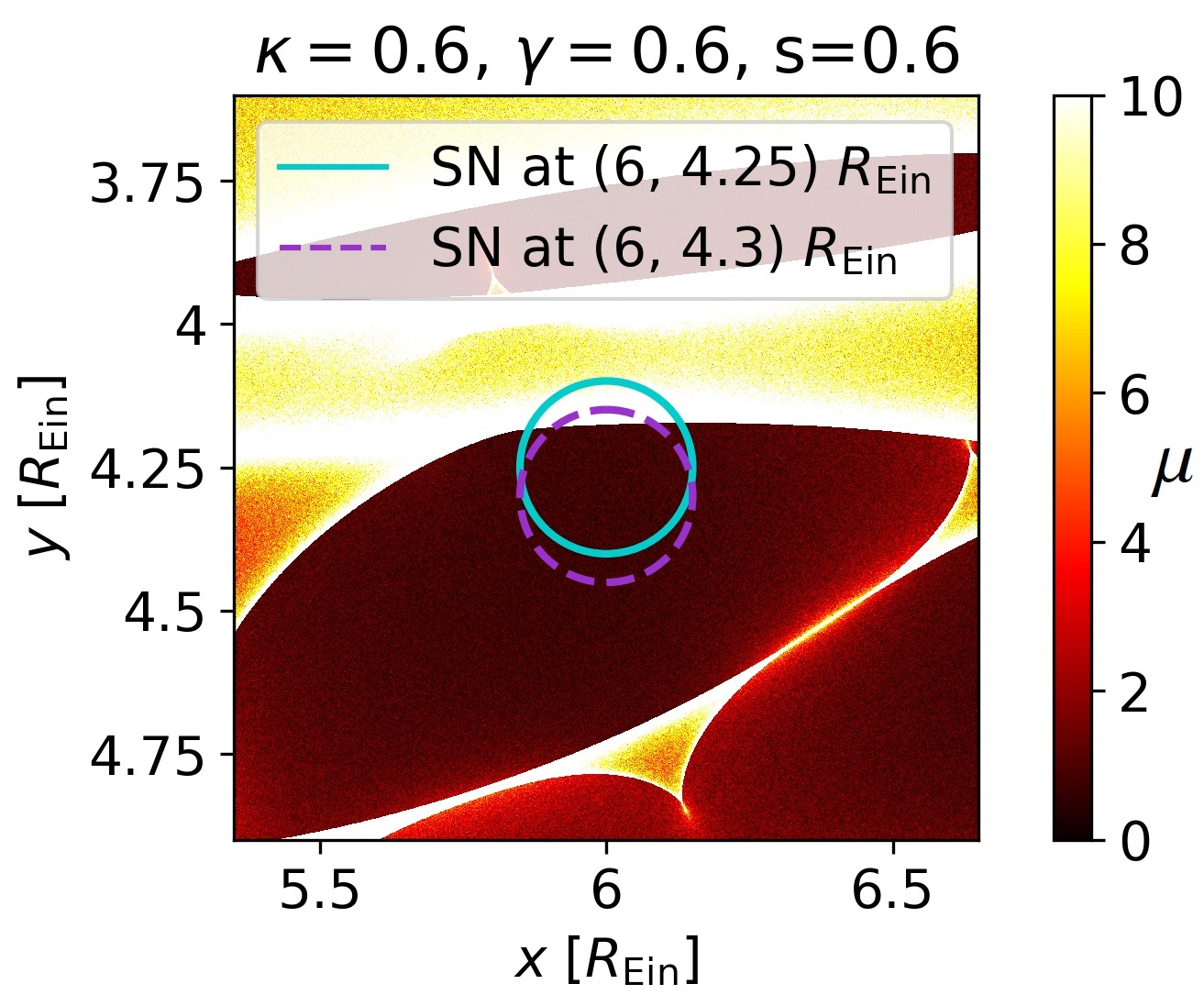}
    \includegraphics[width=0.33\textwidth]{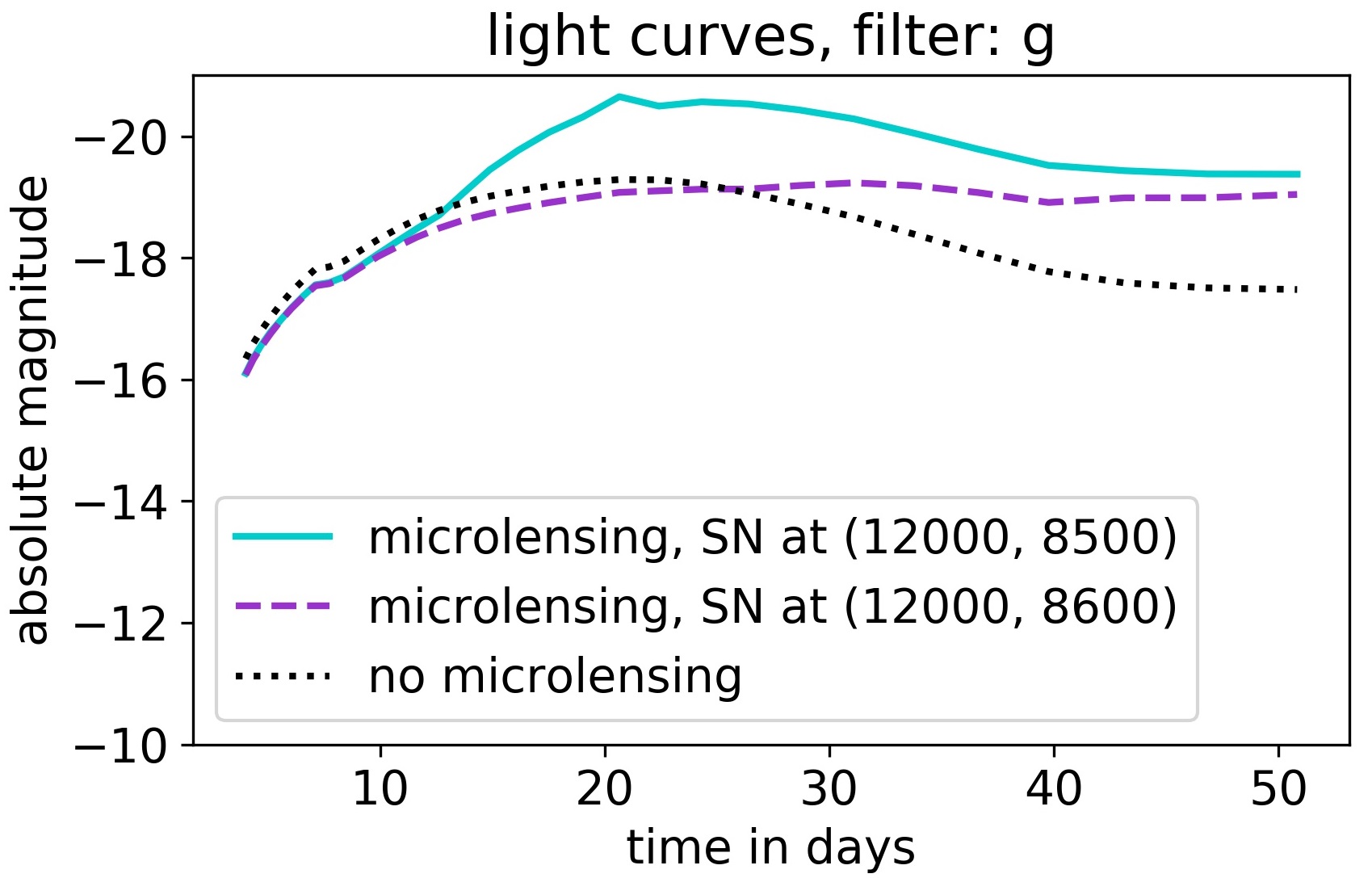}
    \includegraphics[width=0.33\textwidth]{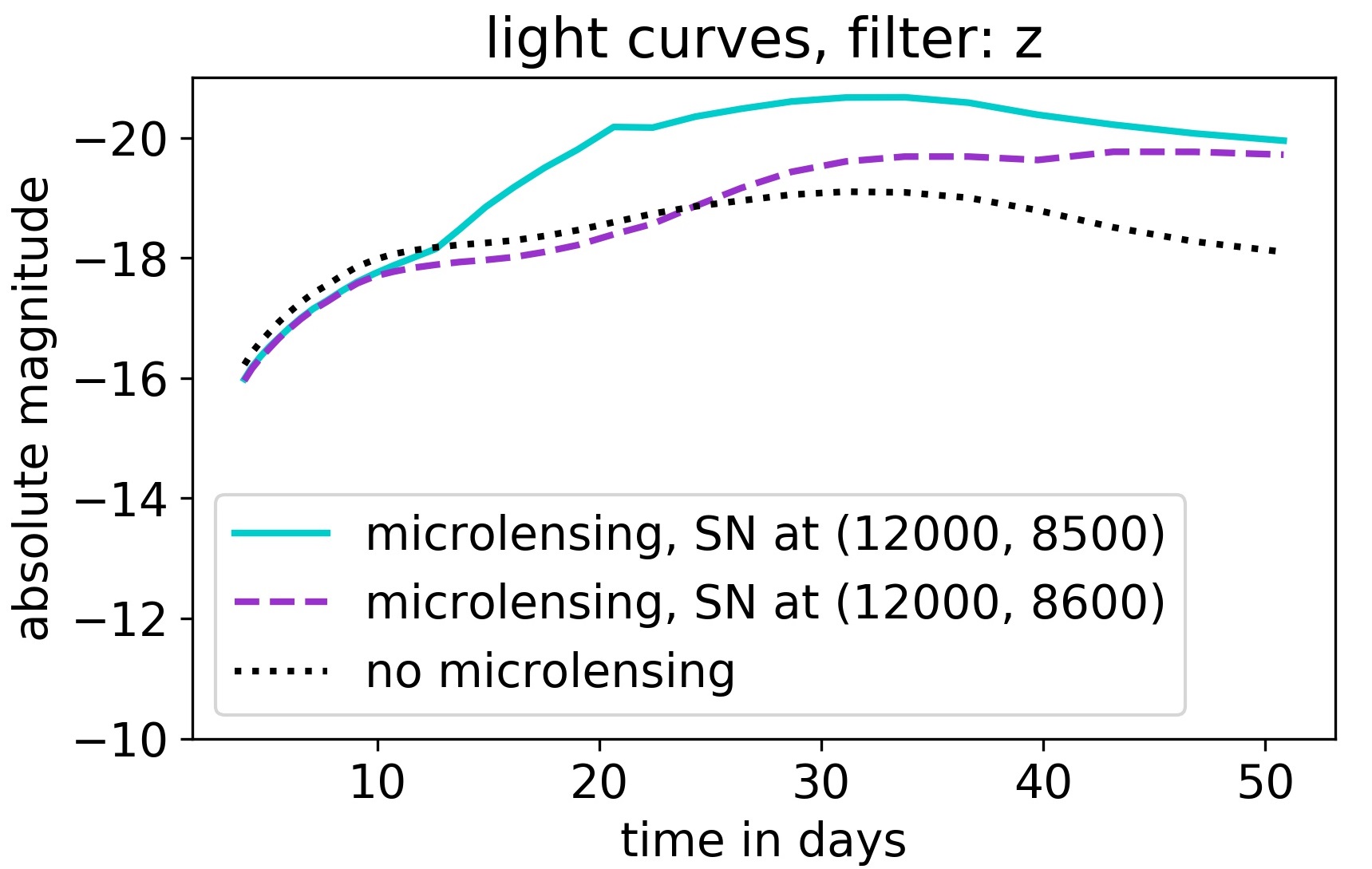}
    \caption{Examples of Type Ia SNe light curves with significant microlensing.  Left panel: a map of microlensing magnifications ($\mu$, indicated by the color bar) on the source plane as a result from stars in the foreground lens galaxy.  The map is for a lensed SN image with convergence $\kappa=0.6$, shear $\gamma=0.6$, and smooth dark matter fraction $s=0.6$.  In this example, the Einstein radius for the mean microlens is $R_{\rm Ein}=7.2\times10^{-3}\,{\rm pc}$.  The two circles in solid cyan and dashed magenta indicate the size of a SN Ia at 21 rest-frame days after explosion. Middle and right panels: microlensed light curves in the g-band (middle panel) and z-band (right panel) corresponding to the SN positions shown in the left panel (solid cyan and dashed magenta).  The intrinsic light curve without microlensing is shown in dotted black.  When the expanding photosphere of the SN crosses a microlensing caustic with high $\mu$, its light curve changes substantially relative to the no-microlensing case. Figure taken from \citet{Huber+2019}. }
    \label{fig:LSNe:cosmo_probe:microlensing_example}
\end{figure*}

Despite the great opportunity presented by having a lensed standard candle, the exploitation of lensed SNe poses an additional challenge: microlensing by stars \citep{DoblerKeeton2006}.
The intrinsic size of a SN is comparable to the Einstein radius of an individual star in the lens galaxy, a case analogous to microlensing of lensed quasars (Vernardos et al.~2023, in prep.). This means that observed magnification is sensitive to the -- essentially unobservable -- positions of stars in the lensing galaxy. Worse, as the supernova expands, the SN atmosphere crosses microlensing caustics. This causes a change in total magnification of the supernova and differential magnification across the atmosphere \citep{Bagherpour+2006, Goldstein+2018, Huber+2019, 2020MNRAS.496.3270M}; this means that the light curves of each image can look quite different even though they are formed from the same background source, as illustrated in Fig.~\ref{fig:LSNe:cosmo_probe:microlensing_example}. This makes it a challenge for measuring accurate time delays between images, though early colour curves can overcome this due to an achromatic phase in the microlensing at early times \citep{Goldstein+2018, Huber+2021} as can spectroscopic observations of absorption lines in the SN atmosphere \citep{Johansson+2021,Bayer+2021}.

The SNe never reach a large enough size for the microlensing effect of multiple stars to average out, so microlensing makes it hard to standardise lensed SNe. The scale of this problem depends in detail on the magnification of the image, the type of the image (time-delay minimum, maximum or saddle) and fraction of density in stars at the location of the image. For high macro-magnification images, the scatter can be as large as 1 magnitude in flux \citep{SchechterWambsganss2002, Yahalomi+2017,2022A&A...662A..34D}, entirely washing out the possibility of using lensed SNe as a standard candle. For images forming further from the Einstein Radius, and particularly for large mass systems where the dark matter fraction is higher at the Einstein radius, the scatter is much smaller and can be less than the intrinsic scatter of a Type Ia SN.

\begin{figure}
    \centering
    \includegraphics[width=\textwidth]{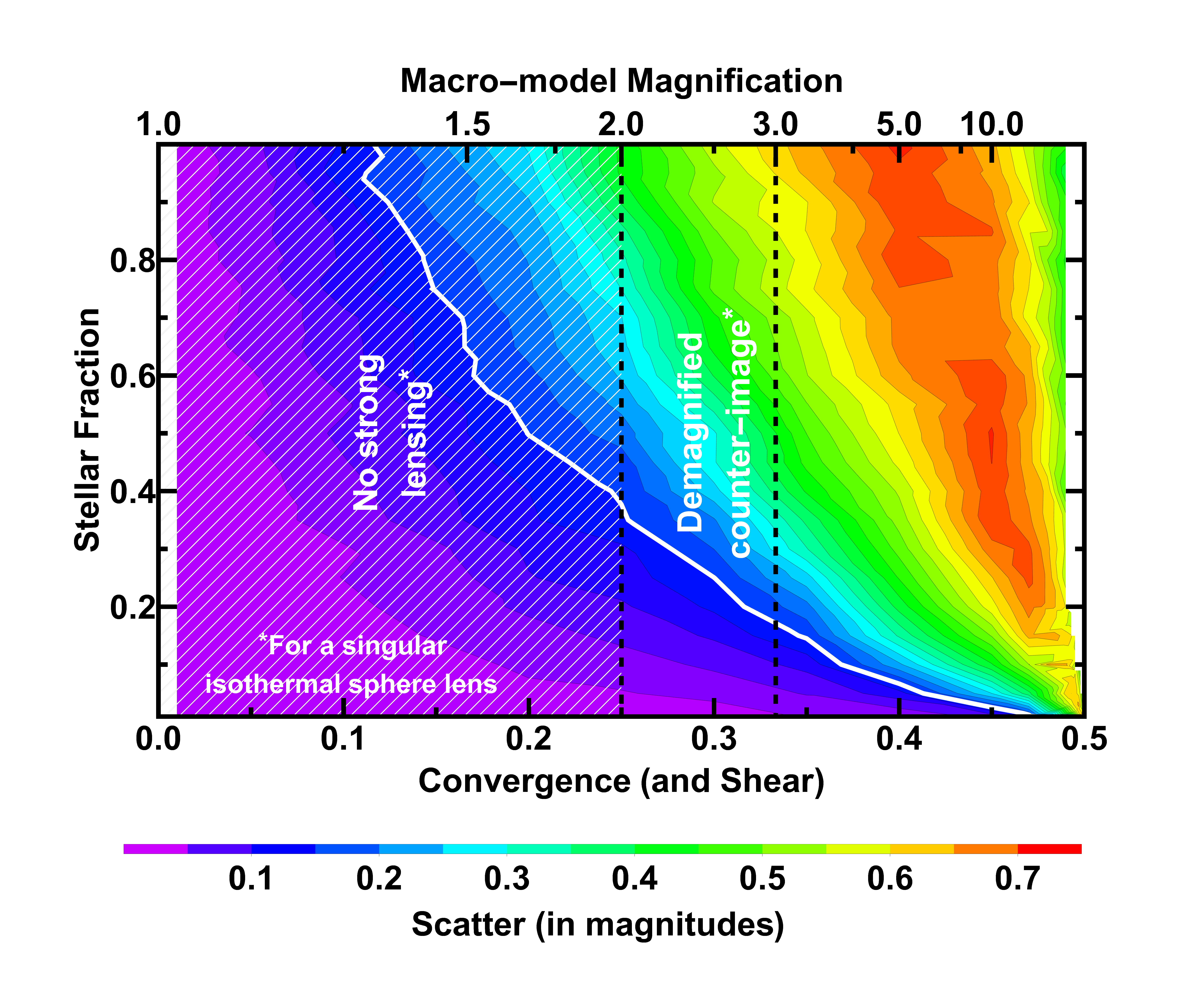}
    \caption{Microlensing induced scatter (indicated by the color bar) for a point source lensed by a Singular Isothermal Sphere macromodel as a function of stellar fraction and macro-magnification.  Scatter is lower for the lensed SN image that is far from the Einstein radius (with lower convergence/shear), but there is a limit since a convergence below 0.25 produces only single images (i.e., no strong lensing) and convergence below $\sim$0.35 produces a faint counterimage (close to lens center) that makes time-delay measurements more challenging.  Image credit: Luke Weisenbach, modified based on \citet{Weisenbach+21}.}
    \label{fig:microscatter}
\end{figure}

\citet{Foxley-Marrable+2018} found that 20 percent of lensed SNe will contain an image with scatter comparable to the intrinsic scatter of a type Ia supernova luminosity, assuming the dark matter fractions of elliptical galaxies derived in \citet{Auger+10} and a Saltpeter IMF. However, these standardisable images only form far beyond the Einstein radius of the lens. This means that the counter images are very close to the centre of the lens and demagnified. Measuring time delays will therefore be challenging, although the faint images arrive second so follow-up with a larger telescope may ameliorate this problem.

The 20 percent standardisable fraction is subject to two key caveats: the stellar initial mass function is assumed to be Salpeter-like and the ellipticity of the lensing mass follows that of the light. 
The assumption of a particular IMF sets the stellar-to-dark matter fraction. Fig.~\ref{fig:microscatter} shows the importance of this assumption: at very high stellar fraction there are no strongly lensed images with minimal microlensing scatter \citep{Weisenbach+21}. With a population of lensed SNe, \citet{Foxley-Marrable+2018} find that observing the amount of microlensing induced scatter can be used to constrain the IMF of the lens galaxy, with strong sensitivity for lenses with Einstein radius between 0.2\arcsec\ and 0.5\arcsec. 
The assumption that mass follows light produces more elliptical lens mass distributions than is seen in cold-dark-matter-only simulations. Increasing the ellipticity of the lens moves the dashed lines in Fig.~\ref{fig:microscatter} to the left. Strongly lensed images can form at lower magnification, where microlensing is less significant. If real lenses are more spherical, then the standardisable fraction will fall.
Therefore, the standardisation of lensed supernovae is subject to somewhat uncertain astrophysics. Exactly what we observe from a large sample of lensed supernovae will inform not just cosmology but also the astrophysics of matter in the strong lensing galaxy.

Despite microlensing distortions on the light curves of SNe and reductions in the number of standardisable SN, lensed SNe still have various advantages over the more conventional lensed quasars for time-delay cosmography.  The drastically varying brightness of lensed SNe enable shorter monitoring campaigns (months) to obtain the light curves and thus more efficient measurements of time delays.  Simulations of microlensed SN light curves with realistic photometric uncertainties showed that time delays can still be recovered accurately and precisely \citep{Huber+2019, PierelRodney2019}, and the dominant source of uncertainty in the time delays is typically photometric uncertainties rather than microlensing distortions \citep{Huber+2021b}.  Another advantage of lensed SNe is that SNe fade after several months, revealing both the lens galaxy and the lensed SN host galaxy more clearly that enable more accurate lens mass modelling \citep[e.g.,][]{Ding+2021}.  In particular, spatially-resolved stellar kinematics of the foreground lens galaxy can be more readily acquired after the SN images fade, and the combination of lensing and kinematic data allows one to break the mass-sheet degeneracy \citep[e.g.,][]{Barnabe+2012, Yildirim+2020, Yildirim+2021, Birrer+2020, Shajib+2023}.  Stellar kinematic maps of the lensed SN host galaxy would further constrain the lens mass distribution \citep{Chirivi+2020}.  We therefore anticipate lensed SNe to be an efficient cosmological probe in the upcoming era of time-domain astronomy when hundreds of such lensed SN events are expected to happen \citep[e.g.,][]{OguriMarshall2010, Quimby+2014, 2019ApJS..243....6G,Wojtak+2019}.

\subsection{Time-delay measurements}
\label{sec:LSNe:cosmo_probe:td_spectro}

The time delays between multiple lensed images are a key ingredient for time-delay cosmography.  The fractional uncertainties in the time delays translate directly to the fractional uncertainties in the $\dt$ and $\dd$ measurements (Equation \ref{eq:cosmo:timedelay}).

\begin{figure*}
    \centering
    \includegraphics[width=\textwidth]{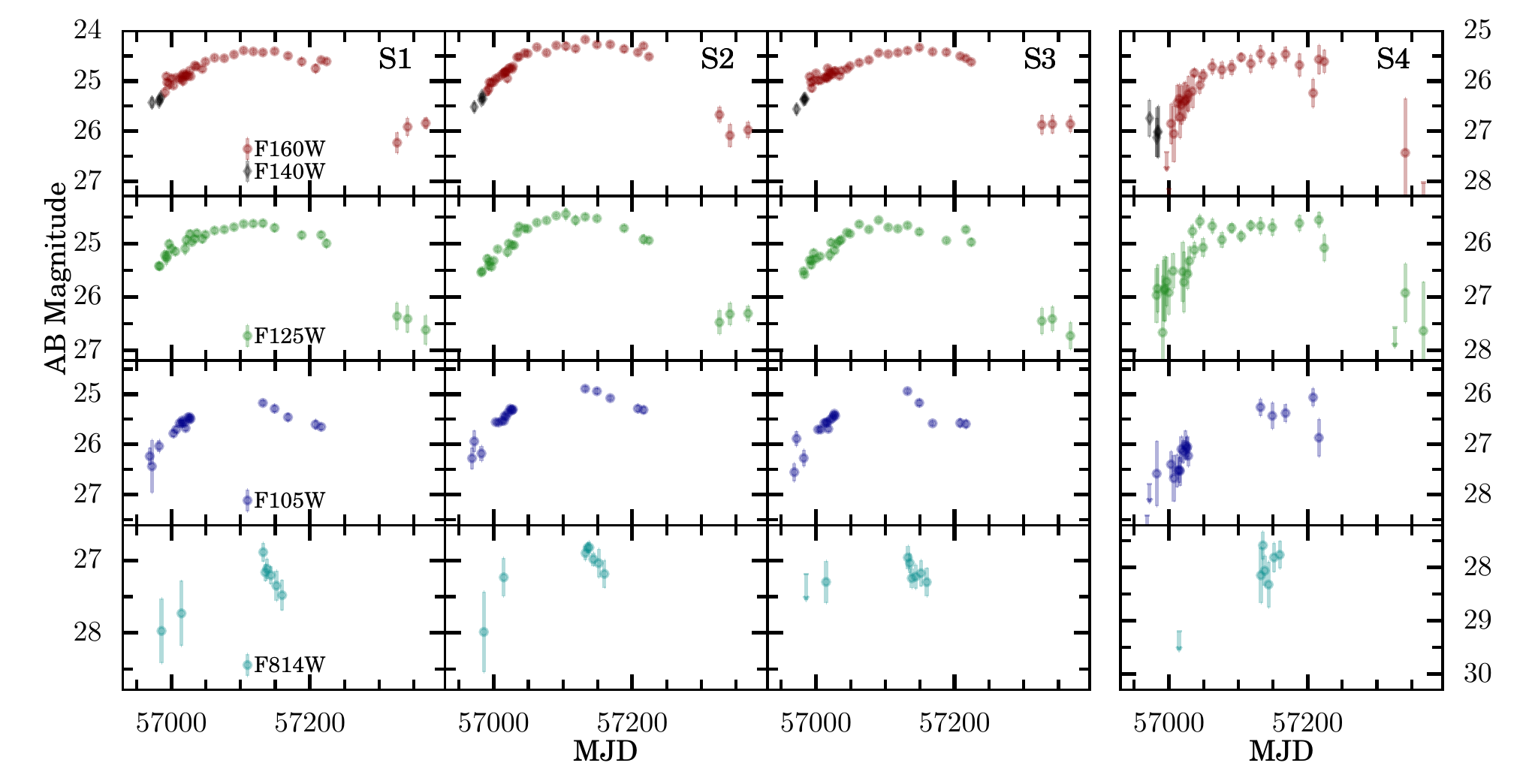}
    \caption{Light curves of SN Refsdal from Hubble Space Telescope imaging in multiple wavelength filters.  Each row consists of the light curves obtained in the specific filter that is indicated in the leftmost panel. Each column shows one of the four SN images, S1-S4 (left to right), that are indicated on the top panels.  Each panel shows the observed AB magnitude as a function of the observer-frame days. Figure taken from \citet{Rodney+2016}. }
    \label{fig:LSNe:cosmo_probe:refsdal_lc}
\end{figure*}

Traditionally, the measurements of time delays involve monitoring the lens system to acquire the light curves of either lensed quasars \citep[e.g.,][]{Fassnacht+2002, Courbin+2018, Millon+2020a, Millon+2020b} or lensed supernovae \citep[e.g.,][]{Rodney+2016, 2020MNRAS.491.2639D}.  Fig.~\ref{fig:LSNe:cosmo_probe:refsdal_lc} shows an example of the light curves of the four images (S1, S2, S3 and S4) of SN Refsdal obtained by \citet{Rodney+2016}.  Curve-shifting techniques that account for microlensing distortions are then applied to these light curves to extract the time delays \citep[e.g.,][]{Tewes+2013a, PierelRodney2019, Millon+2020d}. By making use of the characteristic SN Ia SEDs, \citet{2020MNRAS.491.2639D} used SN Ia template light curves to fit to the HST images of iPTF16geu and measured the 3 independent time delays between the four SN images.  New machine learning approaches such as random forests are also being developed to infer the delays of microlensed SN Ia systems from the light curves \citep[e.g.,][]{Huber+2021b}.

\begin{figure}
    \centering
    \includegraphics[width=0.6\textwidth]{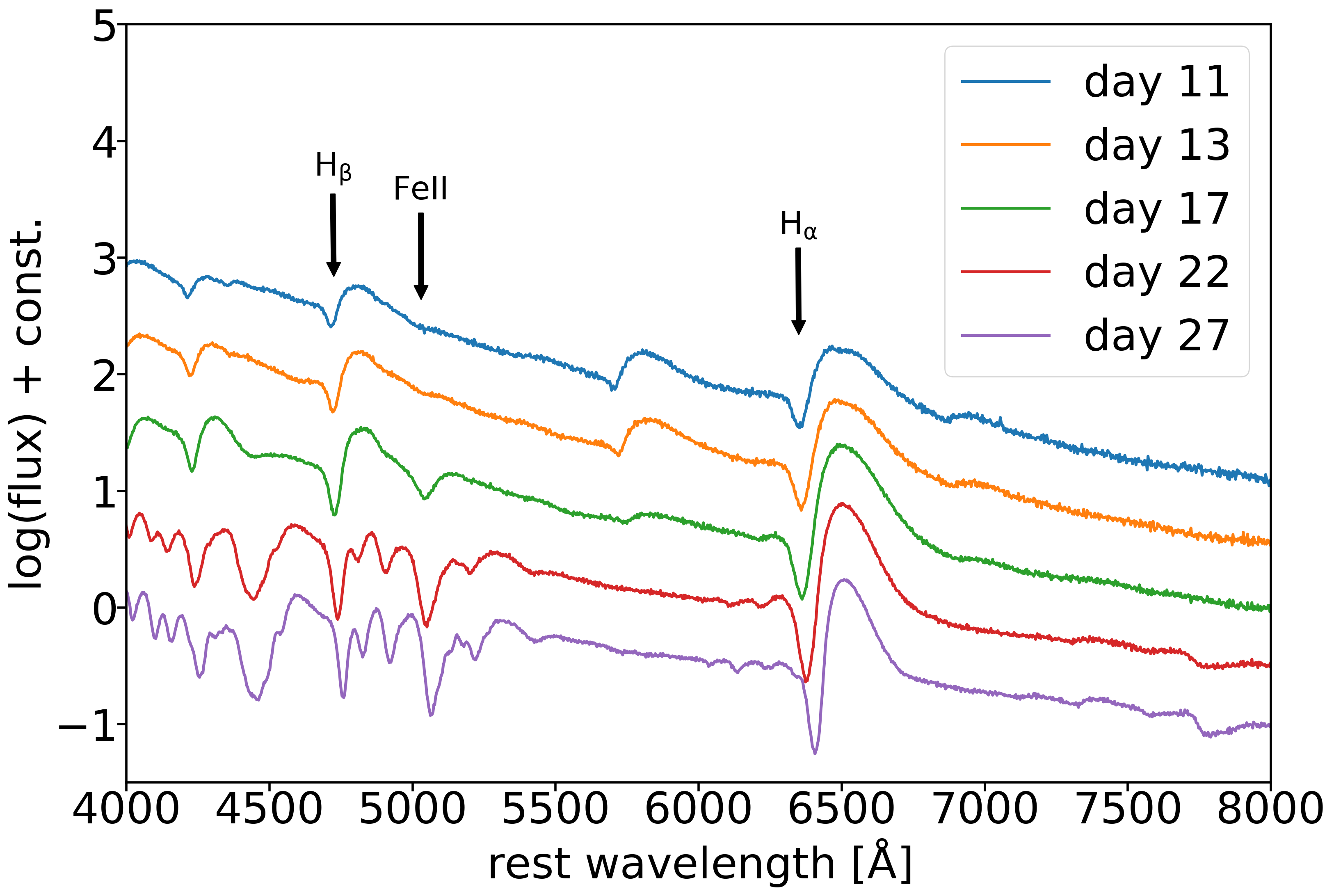}
    \caption{Spectral evolution of core-collapse SN1999em based on the TARDIS simulation from \citet{Vogl+2019,Vogl+2020}. Three prominent spectral features, H$\beta$, FeII and H$\alpha$, are labelled. The SN phases are indicated by their rest-frame days after explosion. As the SN phase increases, the spectral features become stronger and the absorption wavelengths increase.  Such a sequence of spectra of the first-appearing SN image provides the wavelength-phase relation, and a measurements of the absorption wavelength of a trailing SN image therefore provides information of its SN phase and thus its time delay relative to the first-appearing SN image.  With each spectrum of signal-to-noise of 20, \citet{Bayer+2021} showed that the time delays can be measured with uncertainties of $\sim$$2$ days per spectral feature, even after accounting for the effects of microlensing.  Figure taken from \citet{Bayer+2021}. }
    \label{fig:LSNe:cosmo_probe:SN1999em_evol}
\end{figure}

SNe are not only drastically changing their brightness, but their spectra also evolve substantially. The spectroscopic evolution of supernovae offers a new avenue to measure the time delays that are complementary and competitive to the light-curve techniques, as demonstrated by \citet{Johansson+2021} for the case of iPTF16geu. The photospheric velocity in \sneia evolves with time as $v(t) \propto t^{-0.22}$ \citep{PiroNakar2014}, which allows the use of spectral dating to constrain the time delay between images, especially at early times when the velocities of the spectral features is high and the changes from day-to-day are significant. For the most common type of core-collapse supernovae, \sne~IIP, where the light curves rise fast to peak and then reach a plateau phase that could last for hundreds of days, spectral time delays may be the best way forward. 
Fig.~\ref{fig:LSNe:cosmo_probe:SN1999em_evol} shows an example of the spectra of Type IIP SN1999em from the TARDIS simulation \citep{KerzendorfSim2014, Vogl+2019, Vogl+2020} at multiple epochs after explosion \citep{Bayer+2021}.  These epochs (in rest-frame days) are in the plateau phase of this Type IIP SN.  The spectra show prominent absorption lines, especially H$\beta$, FeII and H$\alpha$ that are indicated on the figure, and the absorption wavelength increases as a function of time.  It is precisely such evolutions in the spectral features that allow us to infer the time delays between  two SN images.  

In more detail, a sequence of spectra of the first appearing SN image, say SN image A (such as those shown in Fig.~\ref{fig:LSNe:cosmo_probe:SN1999em_evol}) yields the absorption wavelength of each spectral feature as a function of the SN phase (days after explosion).  When we obtain a single spectra of a trailing SN image, say SN image B, and measure the absorption wavelength of the same spectral feature, we can determine the phase of SN B by using the   relation between the absorption wavelength and phase from SN A.  The determined phase of SN B relative to SN A, together with the known observational times of the epochs, then allows us to compute the time delays between SN images A and B.  Microlensing of SN introduces scatter in the relation between the absorption wavelength and phase, and this effect can be quantified using microlensing maps such as those from the GERLUMPH software \citep{Vernardos+2014a, Vernardos+2014, Vernardos+2015}. Accounting for the effect of microlensing that distorts spectra, \citet{Bayer+2021} have shown that this technique can yield time delays with uncertainties of $\sim$$2$ days per spectral feature with signal-to-noise in the spectra of $\sim$$20$ per wavelength bin of 3\AA\ width.  With multiple spectral features and epochs, the inferred time delays from such spectral approach will be even more precise and accurate.

\subsection{Cosmography with lensed SNe: present cosmological constraints and future forecasts}
\label{sec:LSNe:cosmo_probe:present_future}

\begin{figure}
    \centering
    \includegraphics[width=0.5\textwidth]{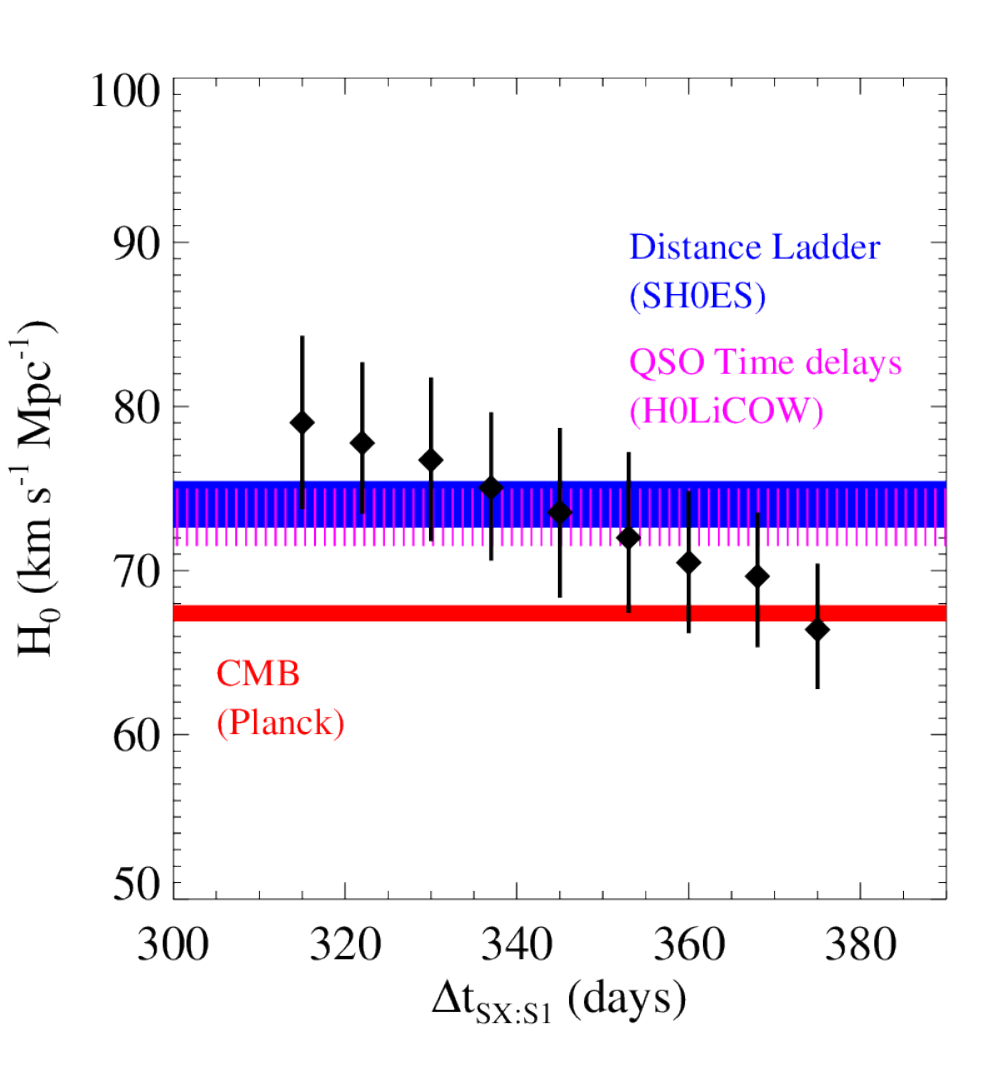}
    \caption{Forecast of $H_0$ from SN Refsdal as a function of the time delay between SN images SX and S1, using the reference mass model of \citet{Grillo+2020}.  The median values of $H_0$ (diamonds) with the 1$\sigma$ uncertainties in flat $\Lambda$CDM are shown for 9 different hypothetical SX-S1 delays, each with an assumed uncertainty of 10 days.  For comparison, the blue, magenta and red bands show, respectively, the 1$\sigma$ credible intervals from SH0ES (\citealt{Riess+2019}), H0LiCOW (\citealt{Wong+2020}) and Planck (\citealt{Planck+2020}).  Figure taken from \citet{Grillo+2020}. }
    \label{fig:LSNe:cosmo_probe:SNRefsdal_H0}
\end{figure}

Of the five known lensed SN systems with spatially resolved SN images, SN Refsdal is the most promising system in delivering a precise and accurate $H_0$ measurement (with $<$10\% uncertainty) in the near future.  As mentioned in Section \ref{sec:LSNe:history}, the measurement of the time delay of image SX relative to the first image S1 is forthcoming, and is expected to have uncertainties of only a few percent (P.~Kelly, private communications).  For a preview of the constraining power on $H_0$ from SN Refsdal, Fig.~\ref{fig:LSNe:cosmo_probe:SNRefsdal_H0} shows a forecast of the inferred $H_0$ for a range of hypothetical time-delay measurements of image SX relative to image S1, assuming the flat $\Lambda$CDM cosmological model \citep{Grillo+2020}.  These are based on cluster mass models of \citet{Grillo+2020} that incorporate all known sources of uncertainties.  We anticipate that a 3\% uncertainty on SX time delay would yield approximately 6\% and 40\% uncertainties (1$\sigma$, including both statistical and systematic) for $H_0$ and $\OM$, respectively.

Various studies indicate that current and future surveys would discover dozens to hundreds of lensed SNe \citep{2002A&A...393...25G,OguriMarshall2010, 2019ApJS..243....6G, Wojtak+2019,2021ApJ...908..190P}.
In particular, \citet{OguriMarshall2010} anticipated thousands of lensed quasars and $\sim$$100$ lensed SNe to be detected in the Rubin Observatory Legacy Survey of Space and Time \citep[LSST;][]{LSST+2009}.
Using a mock sample of $\sim$$1500$ well-observed lenses (consisting of 1476 lensed quasars and 66 lensed SNe) and assuming Planck priors in a flat Universe, \citet{OguriMarshall2010} expected the following 1$\sigma$ uncertainties on cosmological parameters: $\sigma(w_0)=0.15$, $\sigma(w_{\rm a})=0.41$ and $\sigma(h)=0.017$
where $(w_0,w_{\rm a})$ are the time-independent and time-dependent components of the dark energy equation-of-state parameter, and $h$ is $H_0$ in units of $100\,\kmsMpc$.

\begin{figure}
    \centering
    \includegraphics[width=0.7\textwidth]{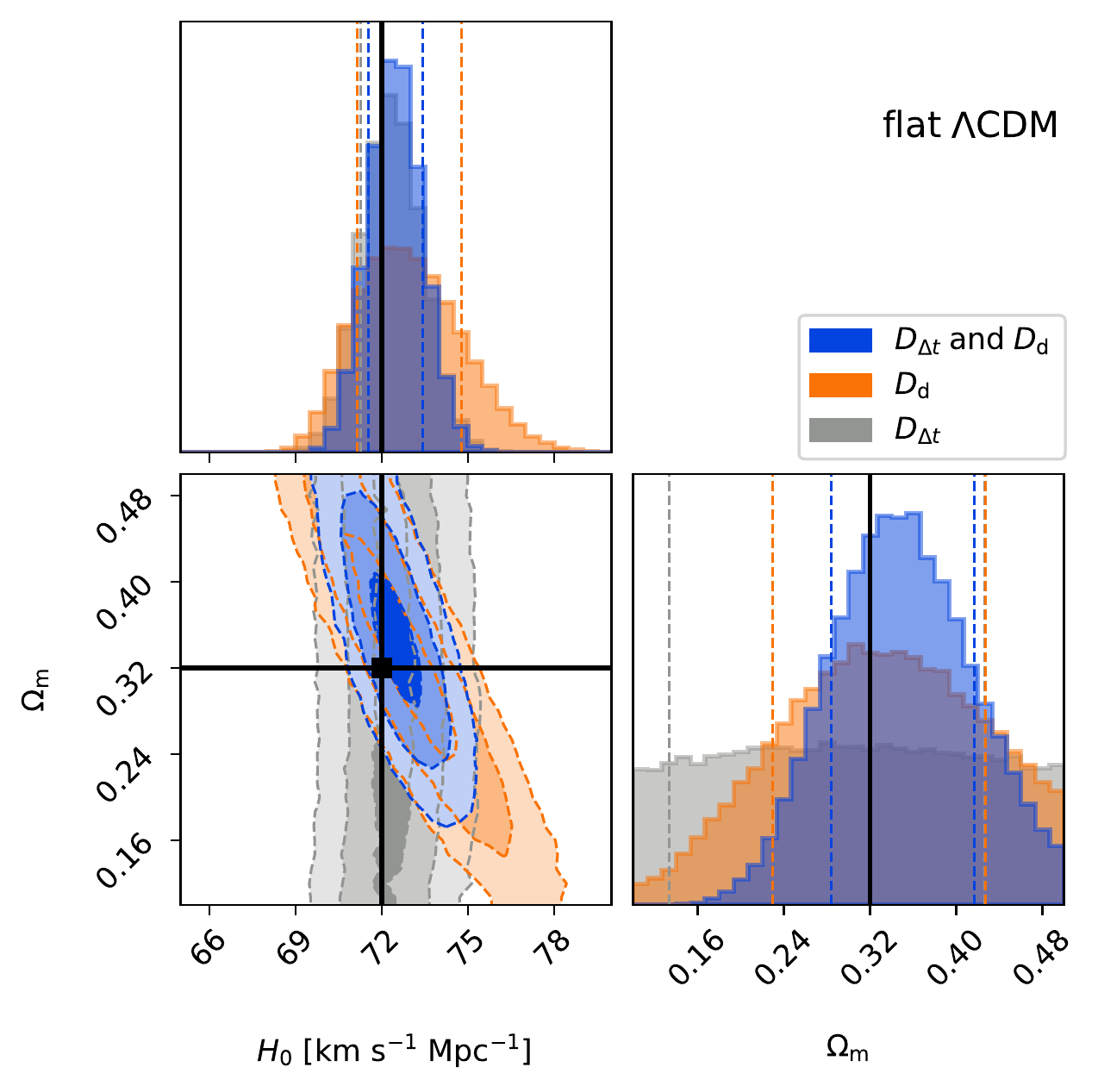}
    \caption{Forecast of constraints on $H_0$ and $\OM$ in flat $\Lambda$CDM from a sample of 20 lensed SNe Ia from LSST with precise and accurate time-delay measurements.  Assuming distance measurements of $\dt$ and $\dd$ with 6.6\% and 5\% uncertainties, respectively, for each lensed SN Ia, this modest sample is expected to yield $H_0$ and $\OM$ with precisions of 1.3\% and 19\%, respectively.   Figure taken from \citet{Suyu+2020}. }
    \label{fig:LSNe:cosmo_probe:LSST_good-delays_cosmo_forecast}
\end{figure}

\citet{Huber+2019} analyzed in detail the number of lensed SNe Ia that could yield time-delay measurements with precisions better than $5\%$ and accuracies better than $1\%$, with realistic microlensed SN Ia light curves and LSST observing strategies.  While LSST is efficient at detecting lensed SNe given its wide survey area and depth, the observing cadence per filter is not rapid enough to map out light curves for precise delay measurements.  Follow-up observations at the cadence of at least one epoch every 2 days would drastically increase the number of lensed SNe with ``good'' time delays (i.e., those delays with uncertainties $<$5\% in terms of precision and $<$1\% in terms of accuracy).  Based on the results of \citet{Huber+2019}, \citet{Suyu+2020} simulated a sample of 20 lensed SNe Ia from LSST that are expected to have good time delays.  Assuming that these systems would have high-resolution imaging and spatially resolved kinematic measurements to break the mass-sheet degeneracy and yield 6.6\% uncertainty on $\dt$ and 5\% uncertainty on $\dd$, this modest sample of 20 lensed SNe Ia could yield constraints on $H_0$ and $\OM$ of 1.3\% and 19\%, respectively, in flat $\Lambda$CDM, as illustrated in Fig.~\ref{fig:LSNe:cosmo_probe:LSST_good-delays_cosmo_forecast}.  In an open $\Lambda$CDM cosmology (i.e., allowing for a spatially curved Universe), a similar constraint on $H_0$ compared to flat $\Lambda$CDM is achievable, while in the flat $w$CDM cosmology (where the dark energy equation-of-state parameter $w$ is allowed to vary), the $H_0$ constraint degrades to 3\% \citep{Suyu+2020}.

The forecast made by \citet{Suyu+2020} did not make use of the standardisable nature of SNe Ia, due to the effects of microlensing and millilensing as mentioned in Section \ref{sec:LSNe:cosmo_probe:candle}.  Nonetheless, \citet{Foxley-Marrable+2018} showed that lensed SNe Ia with asymmetric configurations, i.e., with a SN image located far outside the Einstein radius of the foreground lens, have low microlensing scatter of $\lesssim$0.15\,mag for the outer SN image, which is comparable to the intrinsic dispersion of a typical SN Ia.  These systems are thus standardisable in terms of overcoming microlensing perturbations, and \citet{Foxley-Marrable+2018} estimated that about $\sim$22\% of the $\sim930$ LSST systems predicted by \citet{Goldstein+2018} would be standardisable.  However, \citet{Goldstein+2018} estimated that only $\sim$650 systems would be detected early enough by LSST to deliver reliable time delays. This implies a sample of $\sim$140 lensed SNe Ia that are standardisable and with potential time-delay measurement.

Following the estimates of \citet{Foxley-Marrable+2018} and by making use of 144 standardisable lensed SNe Ia, \citet{Birrer+2021} combined non-lensed SNe Ia with lensed SNe Ia via a Bayesian hierarchical framework to infer the constraints on $H_0$.  The aim is to use the standardisable nature of SNe Ia to break the mass-sheet degeneracy, without using spatially resolved kinematics of the lens galaxy.  Assuming optimistically that the time delays between SN image pairs can be measured with a precision of 2 days for all these systems and that the typical Einstein radius is $\sim$1\arcsec, \citet{Birrer+2021} found that $H_0$ can be measured with an uncertainty of 1.5\%.

The various cosmological forecasts depend sensitively on assumptions of follow-up observations in order to classify the SN type, measure the time delays, acquire high-resolution imaging and spectroscopy.  As described in Section \ref{sec:LSNe:cosmo_probe:candle}, these are necessary for measuring absolute distances to lensed SNe and using them as a cosmological probe. In Section \ref{sec:LSNe:search:followup}, we discuss in more detail the follow-up requirements for lensed SNe.

\section{Astrophysical probe}
\label{sec:LSNe:astro_probe}

Lensed SNe are powerful astrophysical tools to study supernovae and galaxies, in addition to probing cosmology through their time delays. 

In Section \ref{sec:LSNe:astro_probe:SNphysics}, we explain how lensed SNe provide an excellent opportunity to constrain SN progenitors. In Section \ref{sec:LSNe:astro_probe:high-z_spec}, we describe how lensing magnifications enable the acquisition of high-redshift SN spectra that are crucial for SN cosmology.  In Section \ref{sec:LSNe:astro_probe:SN_struct}, we outline the use of microlensing to probe SN structure.   In Section \ref{sec:LSNe:astro_probe:dust}, we show lensed SNe as a probe of dust in the foreground lens galaxy.

\subsection{SN physics and progenitors}
\label{sec:LSNe:astro_probe:SNphysics}

Whilst it is accepted that a Type Ia SN is a thermonuclear detonation of a white dwarf (WD), the cause of the detonations remains uncertain even after decades of study \citep{Maoz2014, Livio2018}. 

The classic progenitor model (often referred to as the single degenerate channel) for a Type Ia SN is a close binary of a white dwarf and a post main sequence star \citep{WhelanIben1973}. As the star expands, it overflows its Roche lobe and its outer layers are accreted onto the WD until the WD reaches the Chandrasekhar mass. At this point, electron degeneracy pressure can no longer support the WD, and gravitational collapse causes thermonuclear detonation of the WD. This model is attractive, since it naturally explains the standardisability of SNe Ia, with the progenitor always having the same mass and near identical composition. However the lack of
evidence for surviving companions suggest that this channel cannot account for all SNe Ia \citep{GonzalezHernandez2012,2015A&A...577A..39L}.

An alternative progenitor model is the double degenerate channel. Here two white dwarfs merge, exceeding the Chandrasekhar mass \citep{IbenandTutukov1984ApJ, Webbink1984ApJ}. This mechanism leaves behind no companion star, but since each pair of white dwarfs will sum to a different mass, there is no obvious reason why this channel would produce a good standard candle.  In addition to these two classic channels of single- and double-degenerate systems, there are other explosion mechanisms that have been explored, including sub-Chandrasekhar explosions \citep[e.g.,][]{Sim+2010}, delayed detonations \citep[e.g.,][]{Ropke+2012}, and double detonations of sub-Chandrasekhar WDs \citep[e.g.,][]{Fink+2007}.

Early observations of SN light curves are critical in constraining the properties of SN progenitor systems (e.g. \citealt{Kasen2010ApJ, Piro2010, Rabinak2011, 2011Natur.480..344N, 2012ApJ...744L..17B, 2015ApJ...799..106G, Piro2016, Noebauer+2017, Kochanek2019, Fausnaugh2019, Yao2019, Miller2020, Bulla2020}). If non-degenerate matter is close to the WD, then it should be shock heated by the explosion, producing excess high energy flux in the first few hours to days of the light curve. In addition to SNe Ia, early observations of core-collapse SNe are also helpful in constraining the properties, such as the sizes, of the progenitor stars \citep[e.g.,][]{Gonzalez-Gaitan+2015,Chen+2022}. Even with the development of wide-field optical surveys, observing these earliest moments of SNe is heavily reliant on chance. Strong lensing gives us the opportunity to predict the precise reappearance time of a SN, such that very early data can be gathered \citep{Suwa2018, Suyu+2020}.

There are multiple challenges to this approach: early detection of the first SN image, possible image demagnification, distortions of SN light curves and spectra by microlensing, and time delay predictions of insufficient precision.  We discuss each of these in turn.

Detection of a lensed SN based on only the appearance of the first image is more difficult, due to possible confusions between a lensed and an unlensed SN without the multiple SN images present. This confusion can be overcome when the lensed SN host galaxy is visible as lensed arc features in addition to the foreground lens galaxy.  

Whilst lensed SNe will be discovered from highly magnified images, images occurring with significant time delay after the bright ``discovery image'' are typically demagnified in the case of two-image systems \citep{FoxleyMarrable2020}.  The situation is better for four-image systems where some of the trailing SN images can be brighter than the discovery SN image.  We expect $\sim2/3$ of the lensed SNe to be two-image systems, and $\sim1/3$ to be four-image systems \citep{OguriMarshall2010}. 

As seen in Fig.~\ref{fig:LSNe:cosmo_probe:microlensing_example} and shown in \citet{Huber+2019}, microlensing distorts the SN light curves and spectra, potentially garbling the information on progenitors.  Encouragingly, \citet{Suyu+2020} have investigated the microlensing effects on four different explosion models of SNe Ia and demonstrated that spectral distortions due to microlensing are $\lesssim1\%$ at the 1$\sigma$ level within 10 rest-frame days after explosions. Therefore, microlensing is expected to have negligible impact on early-phase spectra (within $\sim$10 rest-frame days) for deciphering SN progenitors.  

Finally, time delays cannot be precisely estimated from lens modelling alone.  A 5-10 percent precision is typical from the best lens models \citep[e.g.,][]{Shajib+2019}, and the uncertain value of the Hubble constant given the discrepant $H_0$ measurements \citep[e.g.,][]{Verde+2019, diValentino+2021} adds several more percent of uncertainty in the predicted time delays. For systems with delays $\gtrsim20\,$days that are sufficiently long to catch the trailing SN images from their beginnings (given the time it takes to detect the first SN image and to arrange follow-up observations), a 10\% uncertainty would translate to $\gtrsim2$\,days.  Given this uncertainty, the predicted delay can easily miss the observations within $\sim2$\,days after explosion, the crucially early moments for constraining SN progenitors.

To overcome these challenges, deep and high-cadence (ideally daily) monitorings with imagers after the detection of the first appearing SN image would greatly help.  Such a monitoring would acquire the early-phase light curves of trailing SN images, and spectroscopic observations of the early phases can be triggered as soon as a trailing SN image appears in the monitoring.  Early-phase spectroscopic observations especially in the rest-frame UV are unprecedented and important for constraining SN progenitors \citep{Suyu+2020}.  The time delays of lensed SNe provide a unique and exciting avenue to acquire such spectroscopic observations for studying SN progenitors.

\subsection{High-z SN spectra through gravitational telescopes}
\label{sec:LSNe:astro_probe:high-z_spec}

\begin{figure*}
\centering
\includegraphics[trim= 0 400 0 80, width=0.7\textwidth]{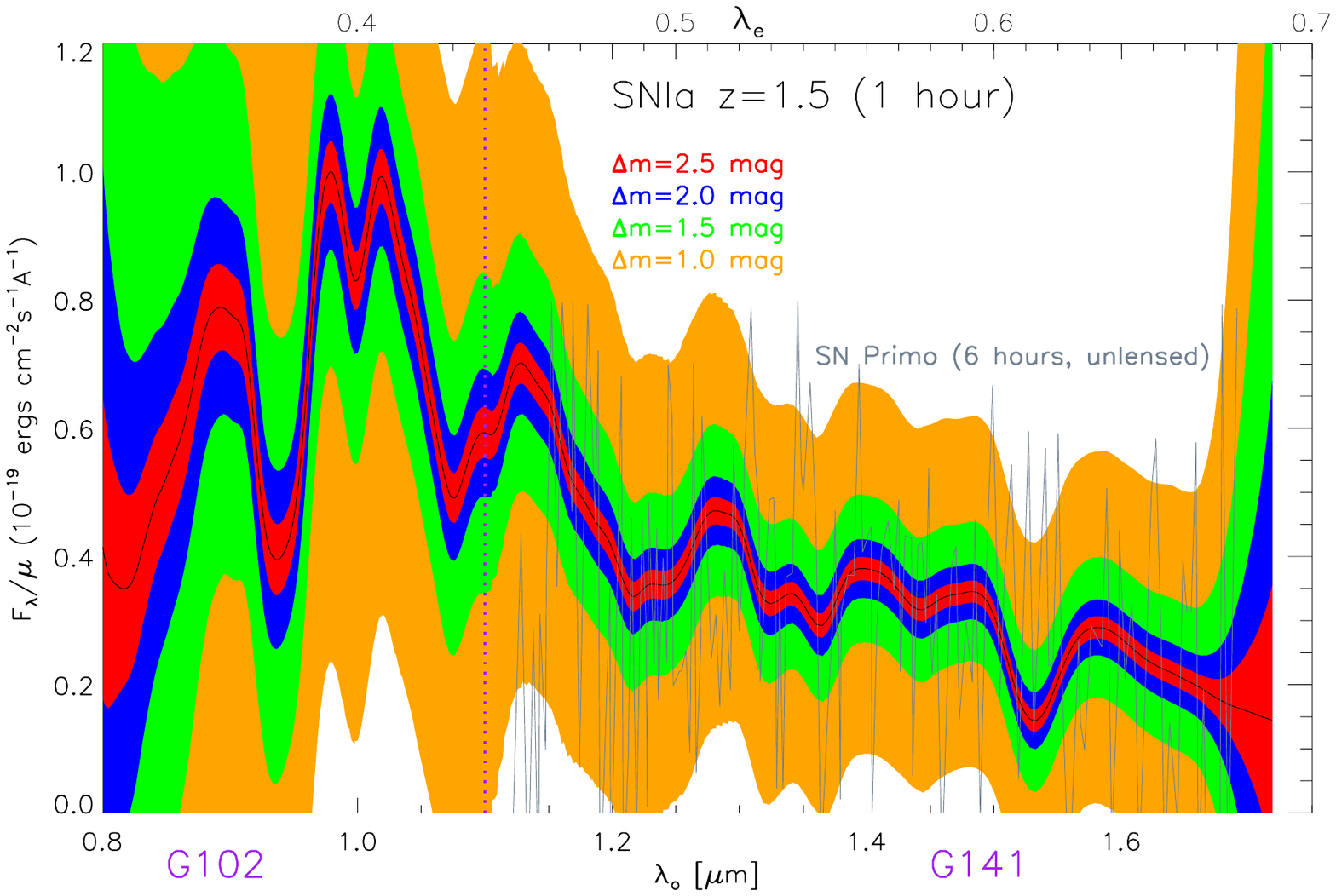}
\caption{Expected signal for HST observations through grisms G102 and G141 of a \snia at $z=1.5$ for different magnifications provided by lensing.
 The shaded regions in $[$orange, green, blue, red$]$ indicate the
  1$\sigma$ uncertainty per pixel in one-hour-long exposures for 
  $\Delta m =[1, 1.5, 2.0, 2.5]$ mag  of  magnification, respectively. The observed spectrum of ``SN Primo'' \citep[][six hours exposure time]{Rodney+2012} is also shown for
  comparison. }
  \label{fig:primo}
\end{figure*}

The precision of \sneia  as distance indicators, and thus their use to
accurately constrain the nature of dark energy, is ultimately limited by our
understanding of progenitor systems, e.g., the potential evolution of the \sn properties, especially their absolute magnitude, over cosmic time.
Thus, confirming the standard-candle nature through spectroscopic comparisons between \sne near and far is essential. 
With current instruments, detailed comparisons beyond $z>1$ are not possible, due to the low SNR. This will  remain challenging in the James Webb Space Telescope (JWST) era. However, multiple spectra of SNe at high redshift are required to check if the progenitor population evolves with redshift. Hence, the magnification of the signal from gravitational lensing could become essential to constrain one of the biggest systematic uncertainties in \snia cosmology \citep{Petrushevska+2017,Johansson+2021}. Fig.~\ref{fig:primo} shows the noisy HST spectrum, based on six hours of observations of SN ``Primo'' at $z=1.55$ observed by \citet{Rodney+2012} compared with the expected signal if observed for shorter time through gravitational telescopes of different strengths, $\Delta {\rm m} = 1, 1.5, 2.0$ and $2.5$ mag, where $\Delta {\rm m}$ is the difference in magnitude coming from lensing magnification. These magnifications are expected to be typical for lensed SNe \citep{2019ApJS..243....6G}.

\subsection{SN structure with microlensing}
\label{sec:LSNe:astro_probe:SN_struct}

When a background source crosses a caustic network produced by microlenses, this allows one to estimate the physical dimensions of the light emitting surface, most robustly its half-light radius \citep{Mortonson+2005,Vernardos+2019}.
For the case of SNe, their expanding ejecta lead to an increasing size that is directly linked to the expansion velocity.
Because the (unobservable) distance from a caustic/high magnification region on the source plane is inversely proportional to changes in brightness due to microlensing, deformed SN light curves, like the ones shown in Fig.~\ref{fig:LSNe:cosmo_probe:microlensing_example}, can be used to constrain the size evolution, and subsequently the expansion velocity. 
In combination with the measurements of the photospheric velocity of the SN ejecta from spectroscopic observations, the SN size evolution can provide information on SN explosion models.
We illustrate the size measurement via microlensing by the following toy model.

Let us assume that the brightness of a lensed SN image at any given time $t$ and wavelength $\lambda$ (ignoring time delays and macromagnification for simplicity) is:
\begin{equation}
\label{eq:toy_model_start}
I(t,\lambda) = \int {\rm d}x \int {\rm d}y \,S(x,y;t,\lambda)\, \mu(x,y),
\end{equation}
where $S$ is the two dimensional intrinsic SN brightness profile, $\mu$ is the microlensing magnification on the source plane, and the integrals are performed over the extent of the source brightness profile.
Assuming, for illustrative purposes, that the intrinsic surface brightness profile ($S$) is a circle of radius $R(t)$ with a known, constant surface brightness $i(t,\lambda)$, we get:
\begin{equation}
\label{eq:toy_model_int}
I(t,\lambda) = i(t,\lambda) \int_0^{2\pi} \int_0^R \mu(r,\theta)\, r\, \mathrm{d}r\, \mathrm{d}\theta = i(t,\lambda) F(R)
\end{equation}
in polar coordinates $(r,\theta)$, where the value of the integral can be written as a radius and time-dependent factor, $F(R)$, and whose probability distribution can be calculated numerically from magnification maps. 
In fact, we can drop the assumption on a uniform profile and perform the convolution in Equation (\ref{eq:toy_model_start}) for different profile shapes.
Therefore, if we know $S(x,y;t,\lambda)$ from a standardised SN type and calculate $F$, then we can measure $R(t)$.
If the SN brightness profile shape varies over time, ratios of $I(t,\lambda)$ can be considered for separations in time $\Delta t$ small enough so that the profile does not change by much. 
Finally, we note that \citet{DoblerKeeton2006} have used a similar approach, the time-weighted light curve derivative, albeit to measure the stellar mass fraction of the lensing galaxy.

\begin{figure*}
\centering
\includegraphics[ width=1.0\textwidth]{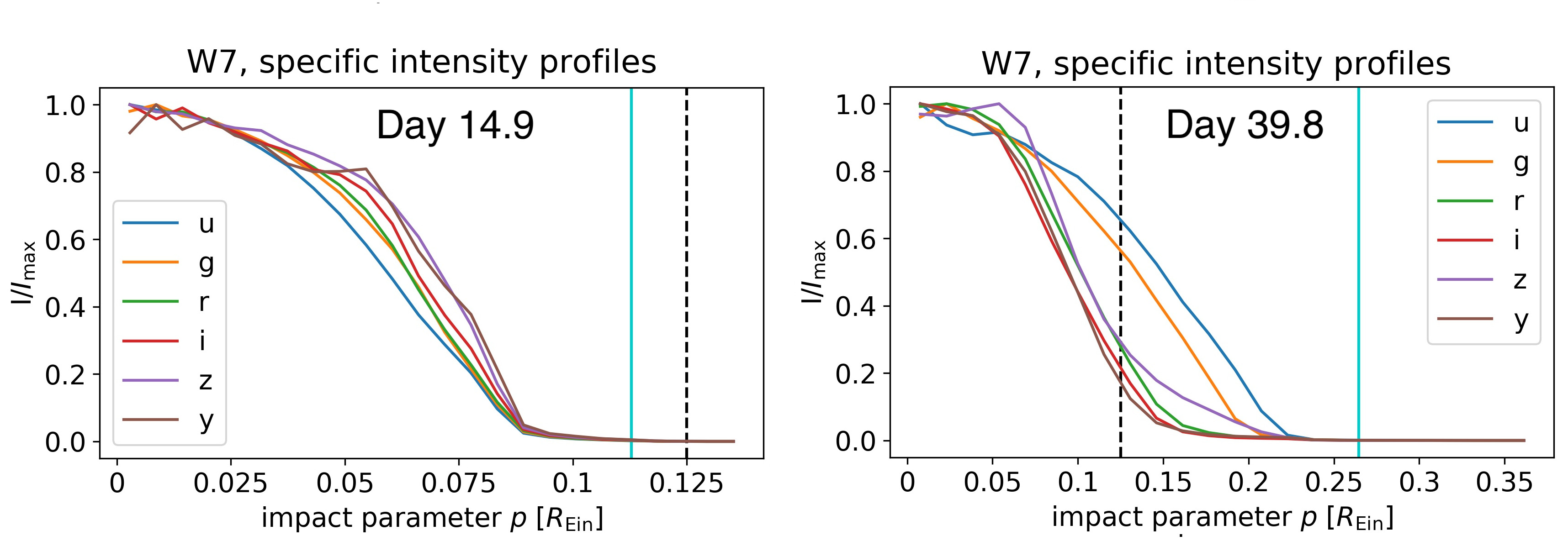}
\caption{Specific intensity profile of the W7 explosion model of SNe Ia.  Left: radial intensity profiles in different filters 14.9 rest-frame days after explosion, in units of the Einstein radius of the microlenses ($R_{\rm Ein}= 2.2\times10^{16}\,{\rm cm}$ for this specific case).  The vertical solid cyan lines indicate the radius that encloses 99.9\% of the total projected specific intensity.  The vertical black dashed lines are random locations of caustics -- effects of microlensing are strong when the specific intensity of the SN crosses a caustic.  Right panel: same as the left panel but for a SN Ia 39.8 rest-frame days after explosion.  Figure extracted and modified from \citet{Huber+2019}.} 
  \label{fig:SNIa_profiles}
\end{figure*}

\begin{figure*}
\centering
\includegraphics[ width=0.35\textwidth]{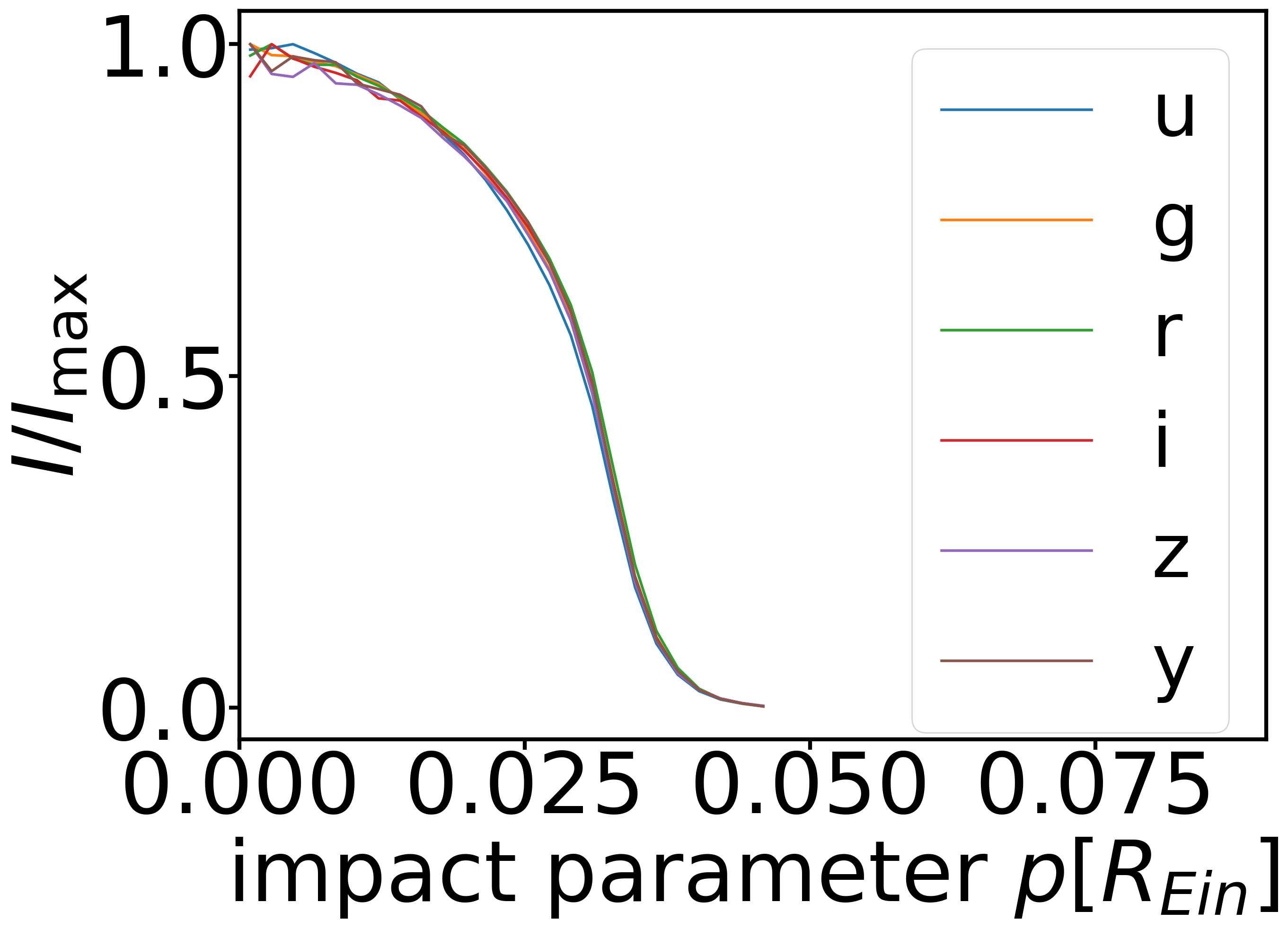}
\includegraphics[ width=0.35\textwidth]{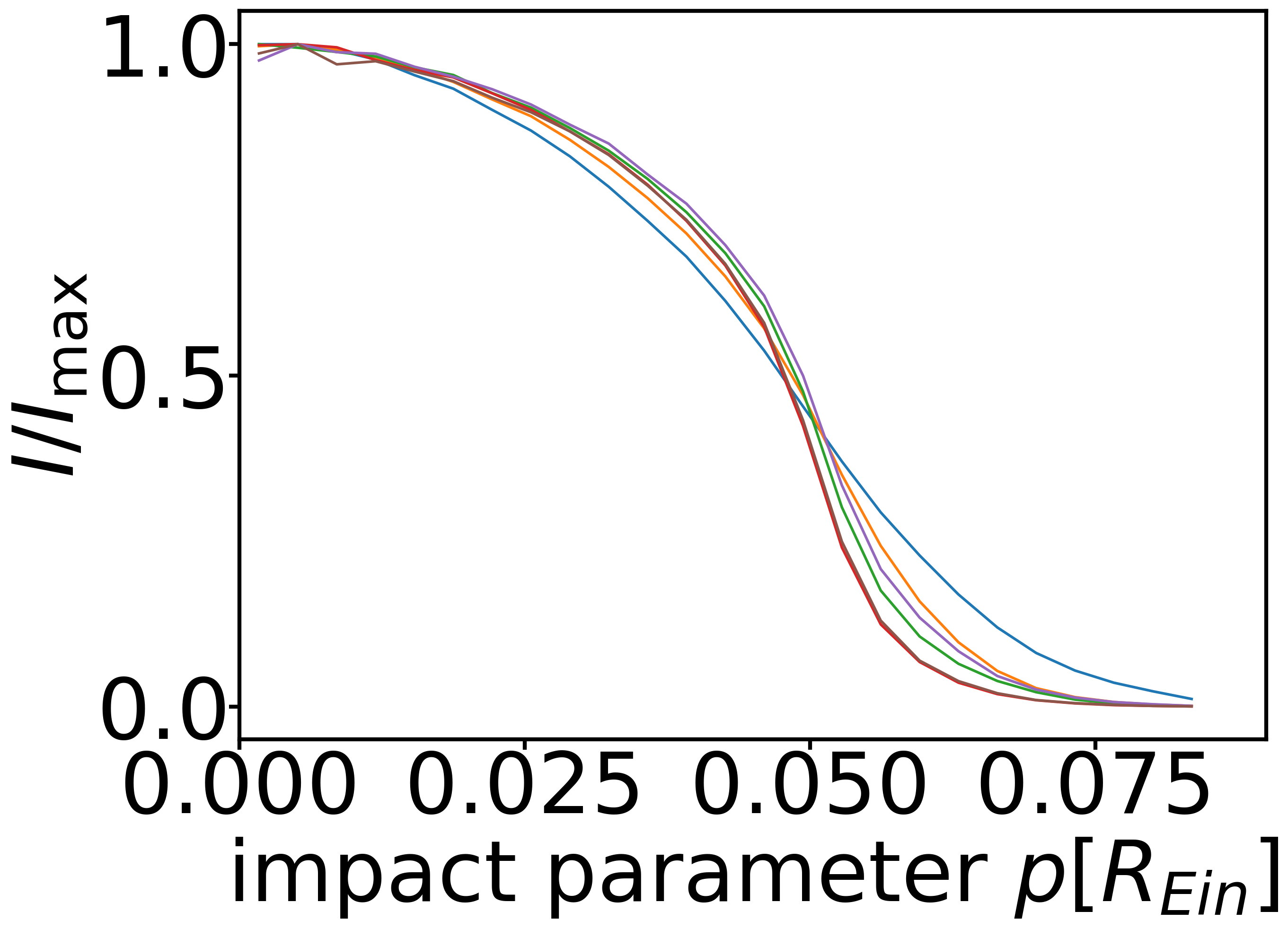}
\caption{Specific intensity profiles from TARDIS \citep{KerzendorfSim2014, Vogl+2019} modelling of SN 1999em, a Type IIP SN \citep{Vogl+2020} projected on the plane of the sky \citep{Bayer+2021}.  Left: radial intensity profiles in different filters 11 rest-frame days after explosion, in units of the Einstein radius of the microlenses ($R_{\rm Ein}= 2.9\times10^{16}\,{\rm cm}$ for this specific case).  Right panel: same as the left panel but for 27 rest-frame days after explosion.  Figure taken from \citet{Bayer+2021}.} 
  \label{fig:SNII_profiles}
\end{figure*}

In an analogous way, the relative size, or more specifically the half-light radius, of the SN ejecta at different wavelengths can be estimated from Equation (\ref{eq:toy_model_start}).
Actually, the shape of the SN profile and/or the expansion velocity may vary in different wavelengths \citep[see Figs.~\ref{fig:SNIa_profiles} and \ref{fig:SNII_profiles};][]{Huber+2019, Bayer+2021}, leading to different half-light radii (i.e., $R = R(t,\lambda)$ instead of $R(t)$).
In the context of our toy model, the brightness ratio between two wavelengths at a given time $t$ can be used to constrain the relative half-light radii:
\begin{equation}
\label{eq:toy_model_lambda_ratio}
\frac{I(t,\lambda_1)}{I(t,\lambda_2)} = \frac{i(t,\lambda_1)}{i(t,\lambda_2)} \frac{F(R_1)}{F(R_2)}.
\end{equation}
Such measurements of radii across wavelengths and time not only provide a check on whether the expansion is homologous (when comparing the radii to the measured velocities from spectra), but also information on the explosion models that alter the dependence of radius on wavelength.
It remains to be seen whether microlensing constraints on explosion models are competitive to the existing constraints based on SN spectral evolution modelling and analysis.

\subsection{Dust properties in distant galaxies}
\label{sec:LSNe:astro_probe:dust}

The different sightlines of multipe images offer unique opportunities to explore the properties of the interstellar medium of the deflecting galaxy, as was first shown for strongly lensed QSOs by \citet{Falco+1999} and in subsequent studies \citep[e.g.,][]{Eliasdottir+2006, 2008A&A...485..403O, Hjorth+2013}. This is important as accurate measurements of e.g., the total-to-selective extinction parameter $R_V \equiv A_{\rm V}/E(B-V)$, a key number for many areas of astronomy, are very hard to establish outside the Milky Way and the Magellanic Clouds. 
The {\it extinction} in the V-band, $A_{\rm V}$, is defined as
\begin{equation}
    A_{\rm V} := m_{\rm V} - m_{\rm V,0},
\end{equation}
where $m_{\rm V}$ is the apparent magnitude with dust extinction and $m_{\rm V,0}$ is the intrinsic magnitude without dust extinction.  The extinction for any other wavelength band is defined in a similar way.  The {\it color excess} (or reddening) between the B and V band is
\begin{eqnarray}
    E(B-V) &:= &A_{\rm B} - A_{\rm V} \\
           & = & (m_{\rm B} - m_{\rm V}) -  (m_{\rm B,0} - m_{\rm V,0}).
\end{eqnarray}
In the case of no dust extinction, then $A_{\rm V} = 0$, $A_{\rm B} = 0$ and $E(B-V)=0$.  For a fixed $R_{\rm V}$ value, higher $A_{\rm V}$ values (or higher $E(B-V)$ values) correspond to more dust along the sight line.   The observed colors in multiple bands of the images are used to infer the differential reddening as the various images travel through different regions of the lensing galaxy (i.e., infer the differences in $A_{\rm V}$ between the multiple SN images). Corrections for differential extinctions in systems of strongly lensed \sneia is crucial to also make use of their standard candle nature to infer the lensing magnification. 

\begin{figure}
 \centering
    \includegraphics[width=0.7\textwidth]{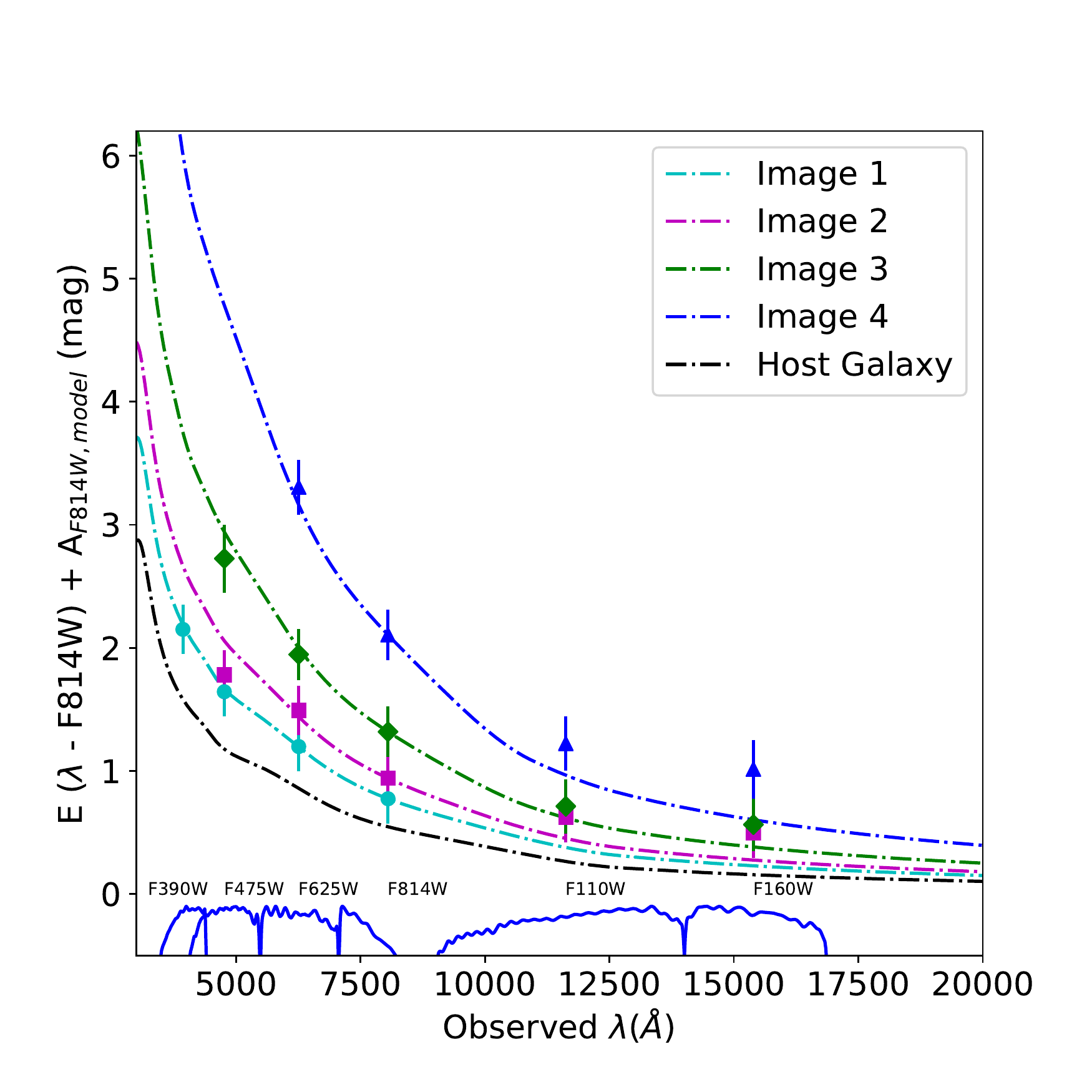}
    \caption{The observed color excess for the resolved images from HST of iPTF16geu as a function of wavelength. The absorption from the host galaxy dust grains is plotted in black. For Image 1 we can see that the host galaxy is the dominant source of extinction, and for images 2, 3, 4 there is a progressively larger contribution from the dust in the lens galaxy, with correspondingly higher values of the color excess. Figure taken from \cite{2020MNRAS.491.2639D}.}
    \label{fig:dust}
\end{figure}

Exploiting the well-known color evolution of \sneia, \citet{2020MNRAS.491.2639D} were able to infer both the extinction in the lensing galaxy for iPTF16geu, along with the common color excess from attenuation in the host galaxy using multi-band images from HST, as shown in Fig.~\ref{fig:dust}. For iPTF16geu, with an impact parameter of about 1 kpc, two of the images (3 and 4) suffered significant extinction and the total magnification inferred from spatially unresolved images was significantly underestimated. Intriguingly, the best fit values of $R_V$ for the lensing galaxy at $z=0.216$ were much lower than what is observed for studies of stars in the Milky-Way \citep{Schlafly+2017}, but similar to what is found in studies of well-measured, highly-reddened  \sneia \citep{2015MNRAS.453.3300A}.      

\section{Searches and Rates}
\label{sec:LSNe:search}

\subsection{Methods}
\label{sec:LSNe:search:methods}

There are multiple ways to find lensed supernovae by making use of their image morphology, magnification, image multiplicity and/or time evolution.  We briefly describe several approaches in this subsection.

\subsubsection{Search through known lensed galaxies}
\label{sec:LSNe:search:methods:known_lens}

A straightforward approach to find lensed SNe is to monitor known lensed galaxies and wait for a SN to explode in one of the lensed galaxies \citep{Shu+2018}.  Using the sample of 128 galaxy-scale strong-lens systems from the Sloan Lens ACS Survey \citep[SLACS;][]{Bolton+2006, Bolton+2008}, the SLACS for the Masses Survey \citep[S4TM;][]{Shu+2017}, and the Baryon Oscillation Spectroscopic Survey Emission-Line Lens Survey \citep[BELLS;][]{Brownstein+2012, Bolton+2012}, \citet{Shu+2018} estimated that the rates of strongly lensed SNe Ia and core-collapse SNe are $1.23 \pm 0.12$ and $10.4 \pm 1.1$ events per year, respectively. 

One can either monitor known lens systems (especially ones with the highest star-formation and SN rates) through dedicated observing programs \citep{CraigEtal21}, or through wide-field imaging surveys such as the ongoing ZTF \citep{Bellm+2019}.  In the latter case, lens systems and lens candidates can be cross matched to transient alerts from ZTF 
through various brokers such as AMPEL \citep[][]{Nordin+2019}, ANTARES \citep{Saha+2014, Narayan+2018, Lee+2020, Matheson+2021} and Lasair \citep[][]{Smith+2019}. Multi-epoch images of lens candidates can also be used to find lensed SNe based on temporal and spatial information of transients occurring near the lens candidates \citep{Sheu+2023}. There are now thousands of confirmed and candidate lens systems from a wide range of lens searches, notably through machine learning approaches in recent years \citep[e.g.,][]{Bolton+2006, Limousin+2009, Gavazzi+2012, Brownstein+2012, Vieira+2013, Canameras+2015, Jacobs+2017, Jacobs+2019, Marshall+2016, More+2016, Sonnenfeld+2018, Sonnenfeld+2020, Petrillo+2019, Canameras+2020, Canameras+2021, Huang+2020, Huang+2021, Savary+2022, Rojas+2022, Shu+2022, Tran+2022}.  However, given the relatively shallow limiting depth of $r=20.6$\,mag (5$\sigma$) of the ZTF survey \citep{Bellm+2019}, individual lensed SN images associated with the lensed galaxies are likely below the detection threshold \citep{OguriMarshall2010, Wojtak+2019}. Nonetheless, the combination of the flux from all lensed SN images can be above the detection threshold, which motivates the next search approach through magnifications.

\subsubsection{Search through lensing magnification}
\label{sec:LSNe:search:methods:magnification}

While targeted searches for lensed \sne in known strongly lensed systems is straightforward, it is also constraining given the limited number of such systems. Since supernovae, especially \sneia, have a known narrow range of intrinsic brightness, it is feasible to identify particularly bright \sne as potential lensing candidates. 

\begin{figure}
    \centering
    \includegraphics[width=0.8\textwidth]{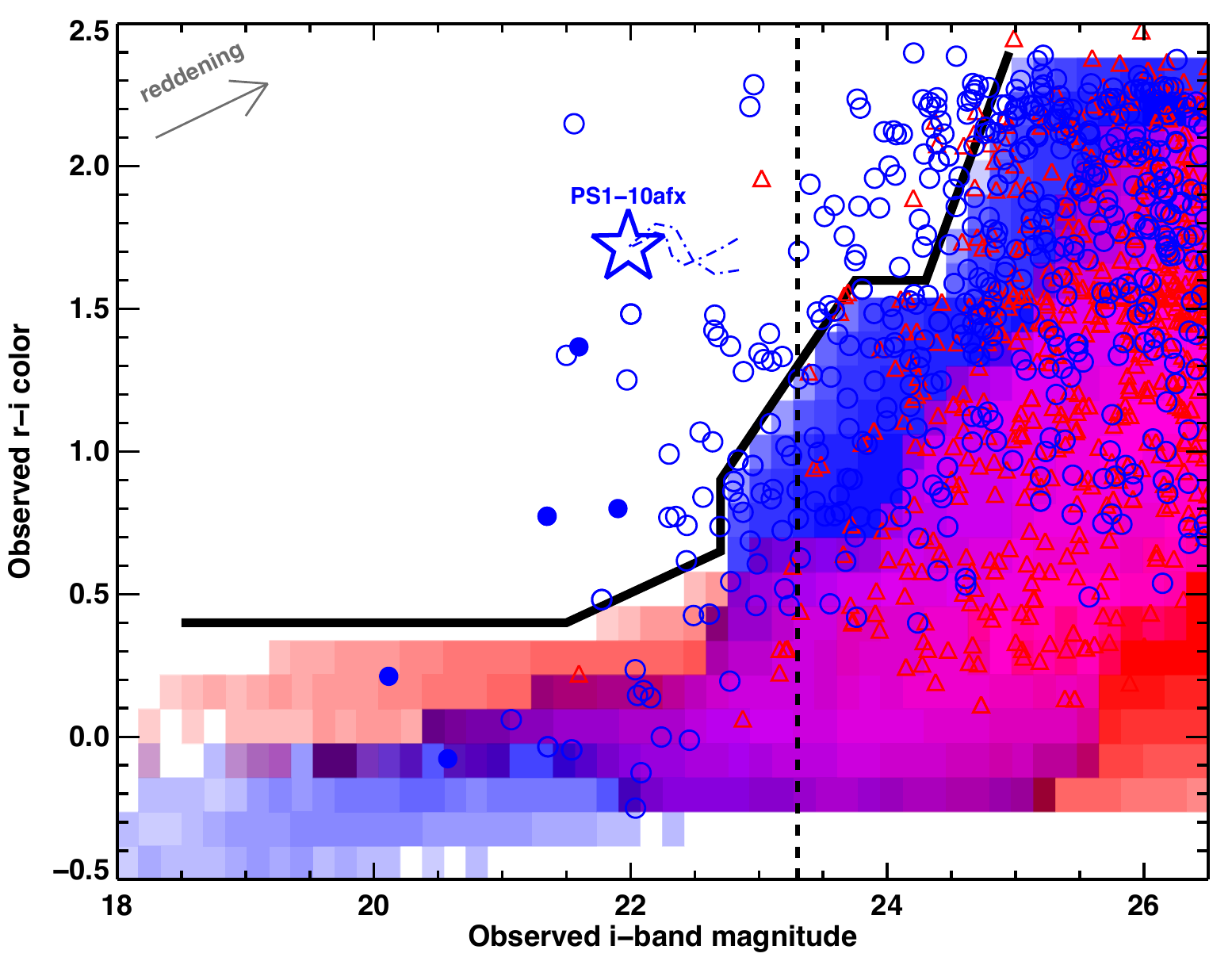}
    \caption{ The expected distribution of unlensed \sne population (\sneia in blue shaded, and core-collapse SNe in red shaded) is enclosed by thick black line showing the red limit. The lensed population predicted by Monte Carlo simulations is shown for \sneia (blue circles) and core-collapse SNe (red triangles). Vertical line marks the single epoch limit for LSST. Figure taken from \citet{Quimby+2014}.}
    \label{fig:colmag}
\end{figure}

\citet{Quimby+2014} proposed a technique to find unresolved lensed images of \sneia based on their high magnification and colors. 
Lensed \sne are expected to come from high redshifts where they will be in larger numbers owing to a larger volume and the lensing probability being higher for distant sources. The lensing magnification will make them appear brighter but their colors will largely be unaffected. Hence, in a color-magnitude diagram, the lensed \sneia will appear redder, at given magnitude, since they will be originating from higher redshifts compared to the unlensed population (see the blue circles above the black solid line in Fig.~\ref{fig:colmag}).

Another technique is based on having some knowledge of the redshifts of the galaxies involved, and it is interesting since it does not require multiple images to be spatially resolved. Thus it can in principle be used to detected arbitrarily compact systems, complementing other techniques, and allowing to probe the entire angular separation distribution of strongly lensed systems.

\begin{figure}
    \centering
    \includegraphics[width=0.8\textwidth]{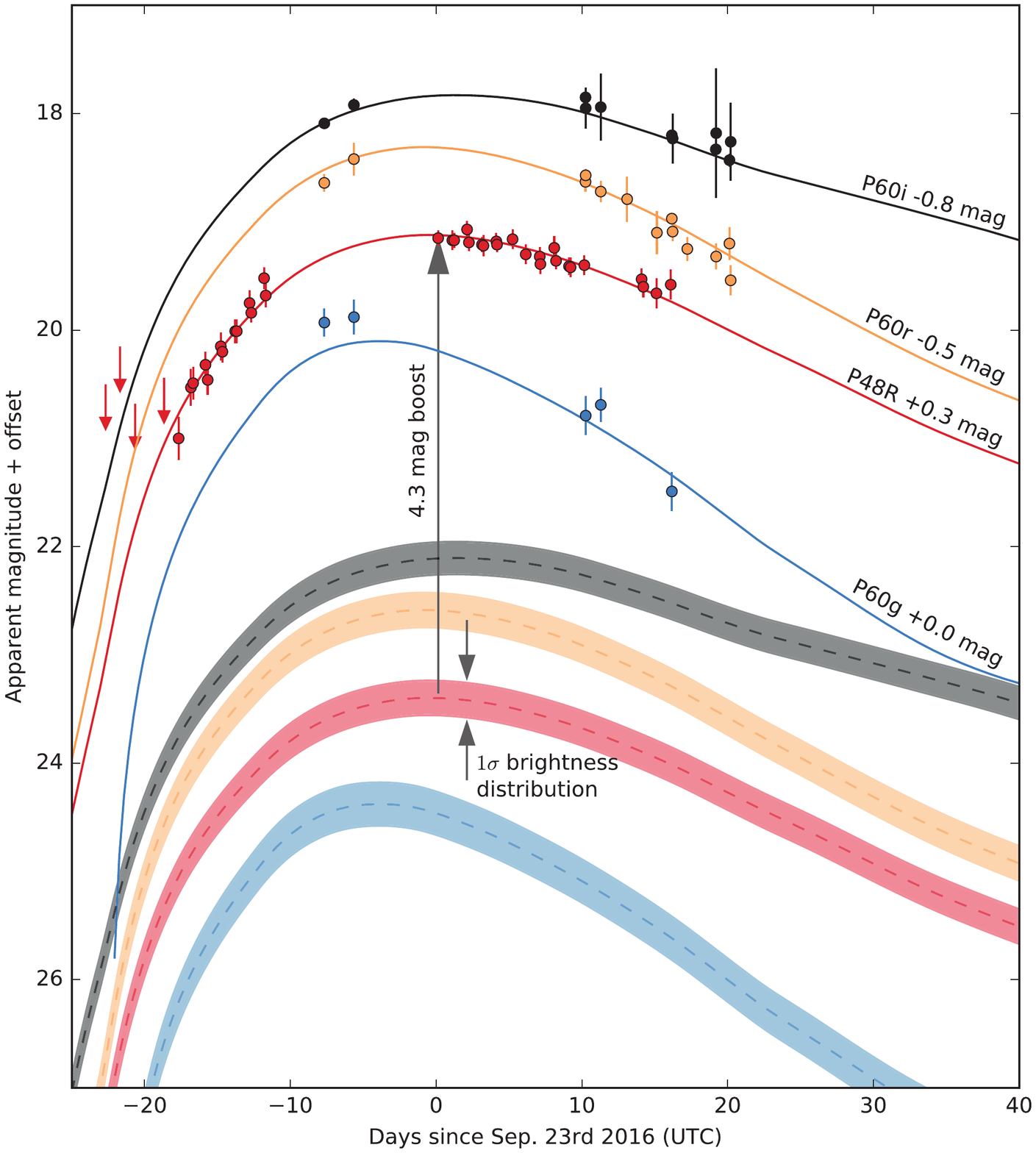}
    \caption{Multicolor light curve of iPTF16geu showing that the unresolved supernova was  4.3 magnitudes (30 standard deviations) brighter than expected for its redshift. The magnitudes are measured with respect to time of maximum light in the R-band at P48 and in the g-, r-, and i-bands with the P60 telescope. The solid lines show the best-fitted \snia model to the data, while the dashed lines indicate the expected light curves at $z = 0.409$ (without lensing). The bands represent the standard deviation of the brightness distribution for \sneia. To fit the observed light curves, a brightness boost from gravitational lensing of 4.3 magnitudes is required. Figure from \citet{2017Sci...356..291G}.}
    \label{fig:16geulc}
\end{figure}

Fig.~\ref{fig:16geulc} from \citet{2017Sci...356..291G} shows a practical realisation of detecting unresolved strongly lensed supernova: iPTF16geu was found to be a 30$\sigma$ outlier when compared with other \sneia at the same redshift. Thus, even with the very modest $\sim2\arcsec$ spatial resolution at the 48-inch telescope at the Palomar Observatory, a strongly lensed system with $\theta_E=0.3\arcsec$ could be identified. The difficulty with this approach is that spectroscopic observations are needed for classification of the transients, along with obtaining reliable redshifts of the galaxies, potentially a daunting task for large surveys with tens or even hundreds of new astrophysical transients found every night. 

\citet{2017ApJ...834L...5G} proposed a methodology to lower the rate of false positives: only considering transients spatially associated with elliptical galaxies, which make up about $\sim 80 \%$ of the galaxy lenses and typically only host \sneia, including the sub-luminous ones.
 Furthermore, thanks to prominent 4000\AA \, breaks, robust photometric redshifts can be computed for elliptical galaxies. 
Hence, if the transient appears to be too bright for being a \snia at the (photometric) redshift of the nearest galaxy, there is a good chance that it is a deflecting galaxy, as opposed to the host. Upon a secure identification, targeted high spatial-resolution imaging can be used to try to resolve the system.

\subsubsection{Search through multiplicity of images}
\label{sec:LSNe:search:methods:multiplicity}

For imaging surveys with faint limiting depth and good angular resolution such as the upcoming LSST, the multiple SN images of a lensed SN can be resolved and detected individually.  \citet{OguriMarshall2010} and \citet{Wojtak+2019} investigated the number of lensed SNe detectable from such an approach.  Specifically, \citet{Wojtak+2019} imposed the following conditions for detecting a lensed SN via its image multiplicity: (1) the maximum image separation between the multiple SN images is between 0.5\arcsec\ and 4\arcsec, where the lower limit is set by the expected seeing of LSST and the upper limit puts focus on galaxy-scale lens systems, (2) the flux ratio between the two images of a two-image (double) system is larger than 0.1, in order to have good contrast and clear identification of the images, and (3) at least three of the four images in a quad system are detected, and both images of a double system are detected.  

\begin{figure}
 \centering
    \includegraphics[width=0.65\textwidth]{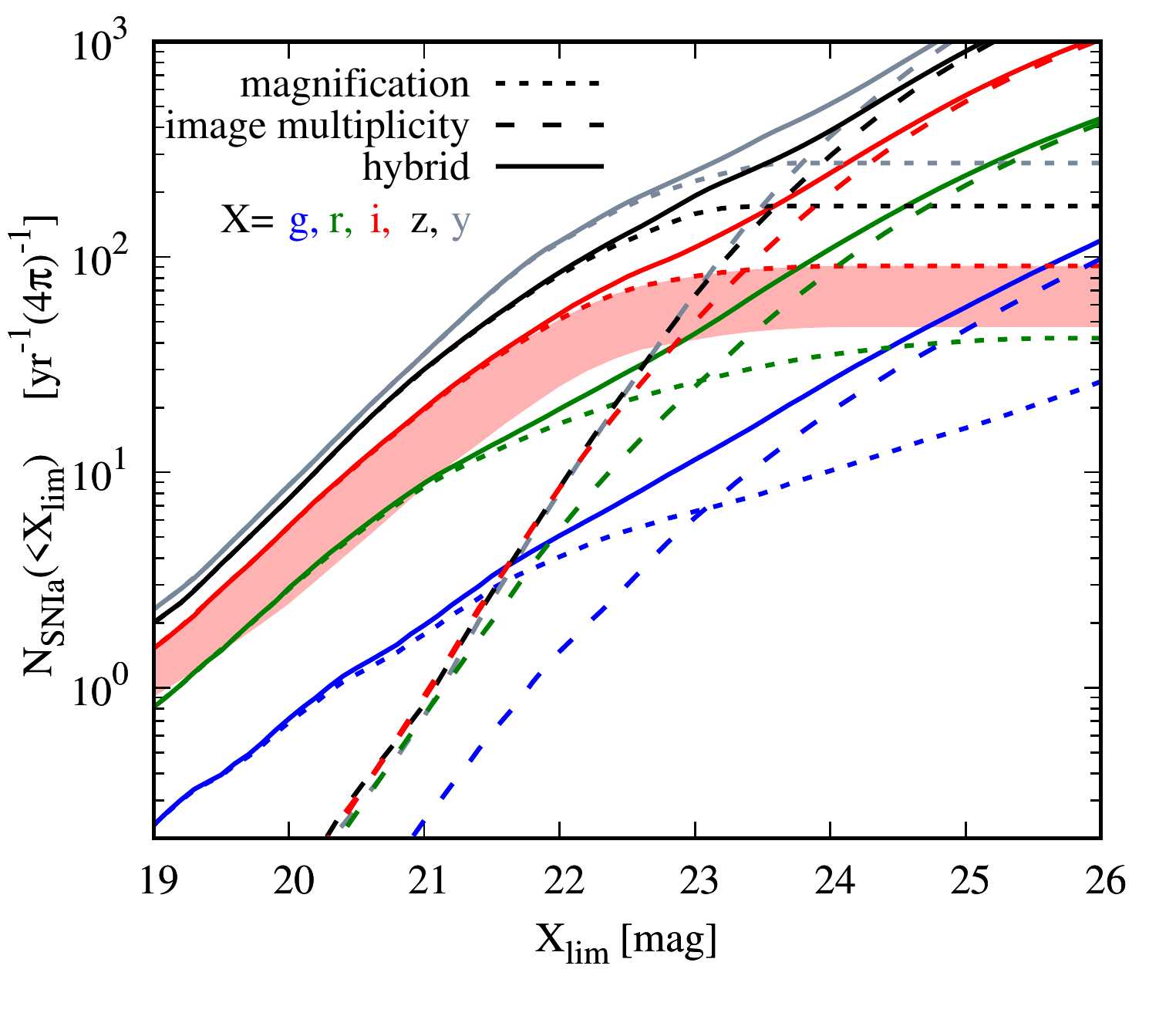}
    \caption{The number of lensed SNe Ia expected per year over the whole sky as a function of the imaging survey limiting magnitude depth.  The different colors correspond to detection in different filters (g, r, i, z or y).  The dotted (dashed) lines show the expected number of lensed SNe Ia detected through the magnification (image multiplicity) approach. The hybrid approach, a combination of magnification and image multiplicity, is indicated by the solid curve.  For shallow image surveys with limiting depth brighter than $\sim$22 such as the ZTF survey, the magnification approach dominates in providing most of the expected lensed SNe.  For deep image surveys with limiting depth fainter than $\sim$23 such as the LSST survey, the image multiplicity approach start to dominate.  Figure taken from \cite{Wojtak+2019}.}
    \label{fig:LSNe:search:mag_vs_mult}
\end{figure}

For surveys with limiting magnitude that are fainter than $\sim$23\,mag in g, r or i-band, the detection via image multiplicity is expected to detect more lensed SN systems than the magnification approach. We illustrate this in Fig.~\ref{fig:LSNe:search:mag_vs_mult} for lensed SNe Ia, and refer to \citet{Wojtak+2019} for other types of lensed SNe that show similar trends.  A combination of the two complementary approaches (magnification and multiplicity), as labelled by ``hybrid'' in Fig.~\ref{fig:LSNe:search:mag_vs_mult}, delivers more lens systems than the individual approaches on their own.  In terms of cosmological applications, the multiplicity approach tends to detect lensed SN systems with longer time delays and larger image separations, which are more suitable for time-delay cosmography.

\subsubsection{Search through spatio-temporal images}
\label{sec:LSNe:search:methods:spatio-temporal}

Most of the lensed SN searches in ZTF are based on the approaches of cross matches to lens systems/candidates (Section \ref{sec:LSNe:search:methods:known_lens}) or magnification (Section \ref{sec:LSNe:search:methods:magnification}) since the angular resolution of ZTF (with 1\arcsec\, pixel sizes) does not resolve the multiple SN images of typical lens systems.  However, in the advent of LSST, there will be high angular-resolution and deep images with at least hundreds of epochs per filter at each sky location. Building upon the image multiplicity method described in Section \ref{sec:LSNe:search:methods:multiplicity}, \citet{KodiRamanah+2021} proposed a machine-learning approach to classify an object based on images from cadenced surveys like the Young Supernova Experiment \citep[YSE;][]{Jones+2021} and LSST.  Machine learning techniques such as convolutional neural networks (CNN) perform well and are efficient at processing large amounts of imaging data for finding gravitational lenses \citep[e.g.,][]{Jacobs+2017, Jacobs+2019, Lanusse+2018, Metcalf+2019, Petrillo+2019, Canameras+2020, Canameras+2021}. The spatio-temporal method of \citet{KodiRamanah+2021} builds upon the success of the CNN and incorporate also time-domain information.

Rather than providing a neural network with the multi-band stacked (static) images of objects for classification, \citet{KodiRamanah+2021} developed a network to take in a temporal series of images.  The network architecture is a CNN that encodes long short-term memory \citep[LSTM, a type of recurrent network;][]{Sherstinsky2020}. The concept is that the temporal series of images will show the multiple SN images appearing at different epochs. Even if the multiple SN images are not well resolved, the centroids of the distribution of light would change as the multiple SN images brighten and dim at different times.  Such change in the features of the object helps the neural network to distinguish lensed SN (with multiple SN images) from non-lensed SN (with single SN image).  Using simulated images of YSE, \citet{KodiRamanah+2021} demonstrated that the spatio-temporal network improves the classification accuracy by $\sim$20\% compared to networks that use static (e.g., single-epoch) images.  The new development spatio-temporal network is very promising for application to the LSST.

\subsection{Expected number of lensed SN events}
\label{sec:LSNe:search:number}

\begin{figure}
    \centering
    \includegraphics[width=0.7\textwidth]{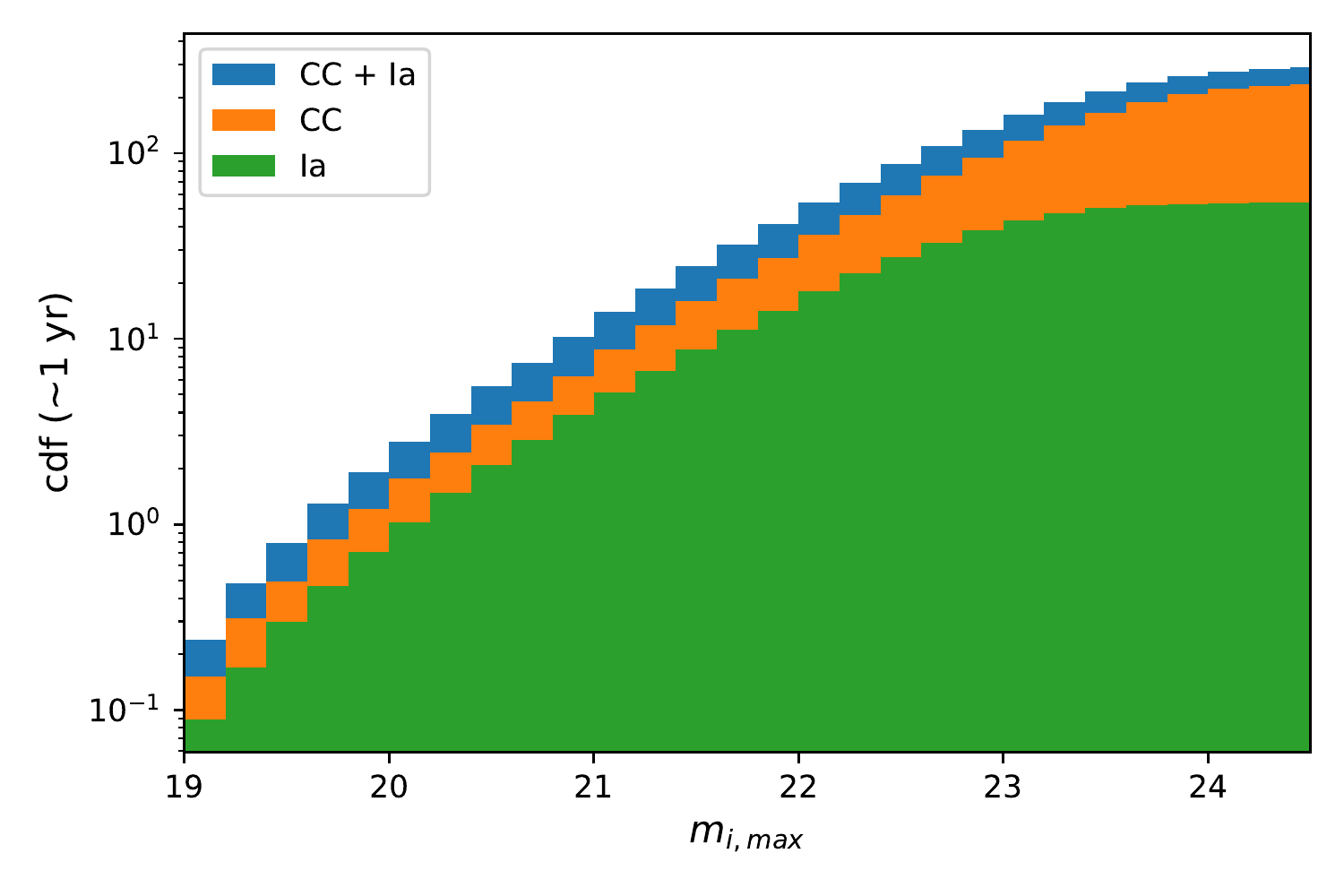}
    \caption{Cumulative number of lensed Type Ia and core-collapse supernovae detections by the LSST survey per year as a function of their peak $i$-band magnitude, as computed in  \citet{2019ApJS..243....6G}. Figure credit: Ana Sagu\'es Carracedo.}
    \label{fig:LSST_rates}
\end{figure}

Strongly lensed transients are very rare. Hence, deep, wide-field time-domain optical and Near-IR surveys have the best chances to find samples of lensed \sne. Among these, the (optical) LSST survey at the Vera C.~Rubin Observatory to see first light in 2024, and the (NIR) Roman satellite, planned for launch some years later offer the best opportunities. In both cases, between several tens to hundreds of strongly lensed supernovae may be expected throughout the multi-year surveys \citep{2002A&A...393...25G, OguriMarshall2010,Quimby+2014,2019ApJS..243....6G,  Wojtak+2019, 2021ApJ...908..190P}. The cumulative number of \sne, Type Ia and core-collapse, expected to be found every year by LSST as a function of the detection threshold is shown in Fig.~\ref{fig:LSST_rates}. A potential concern is the low cadence of the observations, which may require follow-up observations to measure time delays accurately \citep{Huber+2019}.        

\subsection{Expected lens properties}
\label{sec:LSNe:search:lens_prop}

\begin{figure*}
    \centering
    \includegraphics[width=\textwidth]{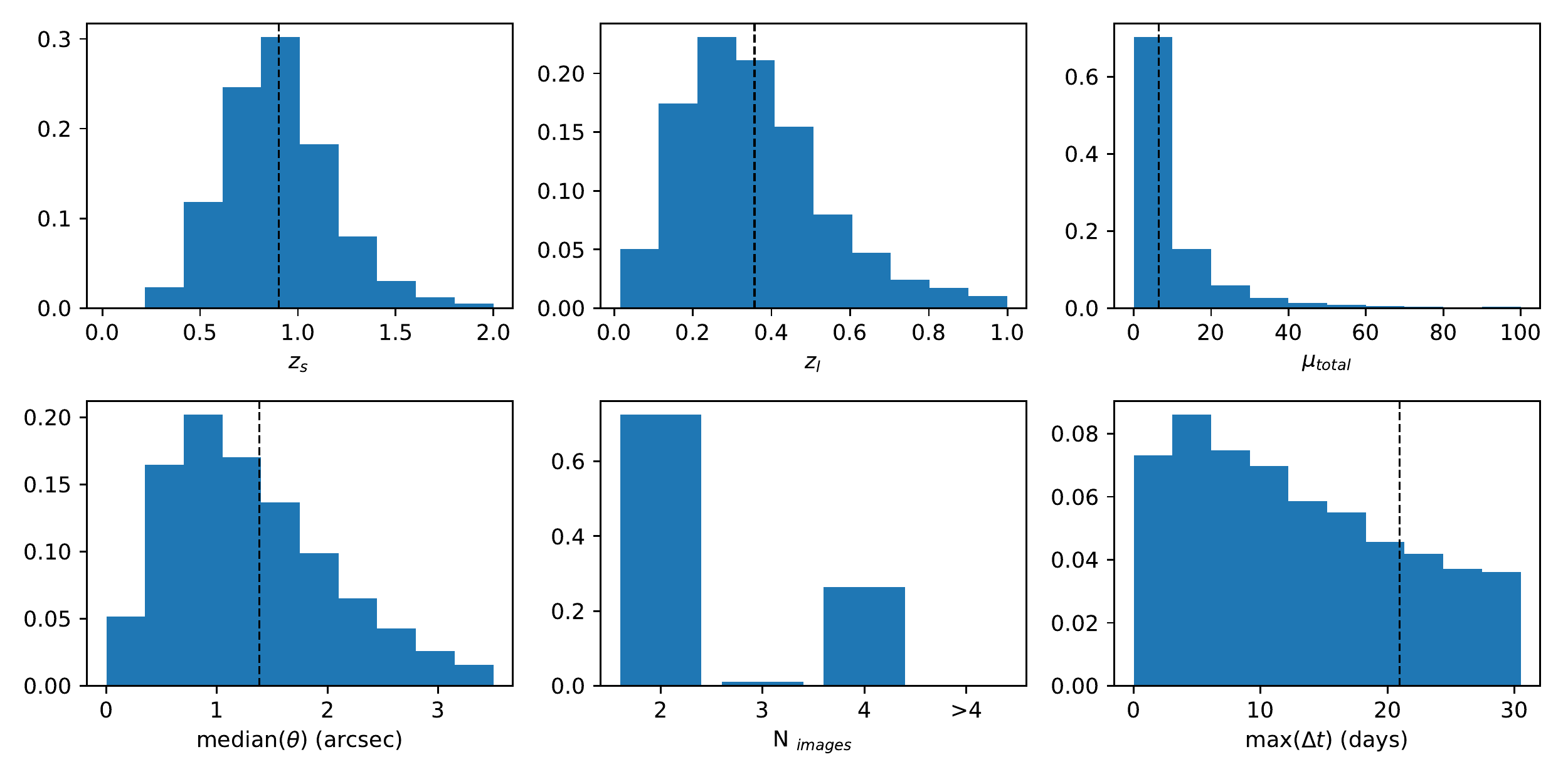}
    \caption{Predicted system configurations for strongly lensed Type Ia supernovae in LSST as computed in  \citet{2019ApJS..243....6G}. Figure credit: Ana Sagu\'es Carracedo.}
    \label{fig:LSST_properties}
\end{figure*}

Fig.~\ref{fig:LSST_properties} shows simulations by \citet{2019ApJS..243....6G} of the distribution of system properties for strongly lensed \sneia in the LSST survey. The typical source redshift is $z_s\sim0.9$ with time delays between 2 and 3 weeks. With a median image separation close to 1$\arcsec$ and the expected good seeing conditions at the Rubin Observatory coupled with the plate-scale of the camera, most events would be spatially resolved.

\subsection{Requirements on follow-up observations}
\label{sec:LSNe:search:followup}

Full scientific exploitation of strongly lensed \sne depends on 1) early transient detection, ideally well before light curve peak; 2) rapid spectroscopic classification of the transient; 3) well-sampled, spatially resolved light curves; and 4) good wavelength coverage, necessary for accurate extinction corrections. From Fig.~\ref{fig:LSST_rates} one can see that the vast majority of the lensed \sne in forthcoming surveys will have (summed) peak magnitudes fainter than 21 mag. Hence, spectroscopic classification requires 8-m class telescopes, in most cases, potentially a major challenge. As argued in Sec.~\ref{sec:LSNe:astro_probe:dust}, an accurate measurement of the image magnifications requires a wide lever arm in wavelength coverage of the spatially resolved images, for at least a handful of epochs along the light curve. The new space telescope, JWST, would be ideally suited for follow-up. Its Near-IR sensitivity is an excellent match to the source redshifts shown in Fig.~\ref{fig:LSST_properties}.  

\section{Summary}
\label{sec:LSNe:summary}

The first discoveries of strongly lensed SNe in recent years are opening a new window of exploration for cosmological and astrophysical studies. In this review, we have provided an overview of the analysis and results from these first lensed SN systems, and the future prospects for this exciting new field.  The main takeaway points are as follows.
\begin{itemize}
    \item There are currently five strongly lensed SN systems with spatially resolved SN images.  Three systems are lensed by galaxy clusters and two lensed by individual galaxies.
    \item Time delays between the multiple SN images of a lensed SN allow direct measurements of (1) the time-delay distance, (2) the angular-diameter distance to the lens in the case where stellar kinematic measurements of the lens galaxy are available, and (3) the luminosity distance in the case of a Type Ia SN that is a standardisable candle, as long as the SN magnifications due to microlensing and millilensing can be accurately accounted for.  These distance measurements provide competitive constraints on cosmological parameters, particularly $H_0$.  
    \item Several new methods to infer the time delays of lensed SNe have been developed in recent years, employing different kinds of data such as light curves, color curves or spectral evolution of SNe.  Time-delay measurements with uncertainties of $\sim$1 day are achievable with real and mock data.  The fractional uncertainty in the delays contribute directly to the fractional uncertainty on the time-delay distance and the lens angular-diameter distance.
    \item Cosmographic constraints, especially the Hubble constant inference, from SN Refsdal are forthcoming.  We expect that a modest sample of 20 lensed SNe Ia from the upcoming LSST to yield $H_0$ with nearly 1\% precision in flat $\Lambda$CDM cosmology.
    \item The time delays of lensed SNe provide a unique opportunity to acquire very early-phase observations of SNe, especially in the rest-frame UV, and probe SN progenitors.
    \item Lensing magnifications allow the acquisition of spectra of SN above $z>1$, that are crucial to reduce systematic uncertainties in SN Ia cosmology.
    \item Microlensing of SNe provides an avenue to constrain the sizes of the SN at different wavelengths.
    \item The dust properties in the (foreground lens) galaxies can be measured through the multiple SN sight lines in lensed SN systems.
    \item Various methods have been developed to search for lensed SNe, through e.g., the monitoring of known lens systems, magnification, image multiplicity, and spatio-temporal evolution in the imaging.  Depending on the selection criteria, we expect at least dozens of lensed SNe to explode in the upcoming LSST, with their properties summarized in Section \ref{sec:LSNe:search:lens_prop}.

\end{itemize}

We are entering a new era of lensed SNe with the upcoming wide-field cadenced imaging surveys.  Rapid follow-up observations including spectroscopic typing and light curve monitoring, will be crucial and necessary to make the best use of such events for cosmological and astrophysical studies.

\begin{acknowledgements}
We thank Luke Weisenbach for producing Fig.~\ref{fig:microscatter}, and Raoul Ca\~nameras and Paul Schechter for discussions and feedback on the manuscript.  
We are grateful to the International Space Science Institute (ISSI) in Bern for the hospitality and the stimulating workshop on $``$Strong Gravitational Lensing$"$.
SHS thanks the Max Planck Society for support through the Max
Planck Research Group and the Max Planck Fellowship.
This project has received funding from the European Research Council (ERC)
under the European Union's Horizon 2020 research and innovation
programme (LENSNOVA: grant agreement No 771776).
This research is supported in part by the Excellence Cluster ORIGINS which is funded by the Deutsche Forschungsgemeinschaft (DFG, German Research Foundation) under Germany's Excellence Strategy -- EXC-2094 -- 390783311.
AG acknowledges support from the Swedish National Space Board and {\em Vetenskapsr\aa det},  the Swedish Research Council.
TEC is funded by a Royal Society University Research Fellowship.
GV has received funding from the European Union's Horizon 2020 research and innovation programme under the Marie Sklodovska-Curie grant agreement No 897124.
GV's research was made possible by the generosity of Eric and Wendy Schmidt by recommendation of the Schmidt Futures program.

\end{acknowledgements}

\bibliographystyle{aps-nameyear}      
\bibliography{main}                
\nocite{*}

\end{document}